\documentclass[11pt,reqno]{amsart}

\usepackage{amssymb,amsmath,amsthm,amscd,latexsym,amsfonts}
\usepackage{mathtools}
\usepackage[T1]{fontenc}
\usepackage{graphicx,epstopdf}
\usepackage{graphicx}
\usepackage{xcolor}
\usepackage{cite}
\usepackage{float}
\usepackage[a4paper,top=3.1cm, bottom=3.1cm, left=3.3cm, right=3.3cm]{geometry}

\newtheorem{thm}{Theorem}
\newtheorem{defn}{Definition}
\newtheorem{lemma}{Lemma}
\newtheorem{pro}{Proposition}
\newtheorem{rk}{Remark}

\numberwithin{equation}{section} 

\newcommand{\B}{{\mathcal B}}
\usepackage[mathscr]{eucal}
\usepackage{mathrsfs}

\newcommand{\bea}{\begin{eqnarray}}
	\newcommand{\eea}{\end{eqnarray}}

\def\l{\lambda}

\begin{document}
	\title [Non-Linear Generalization of the DLR Equations]
	{Non-Linear Generalization of the DLR Equations: 
$q$-Specifications and 
$q$-Equilibrium Measures}
	
	\author{F.H.Haydarov, B.A.Omirov, U.A. Rozikov}
	\address{%
  F.H. Haydarov$^{a,b}$  
  \begin{itemize}
 \item[] $^a$ V.I.Romanovskiy Institute of Mathematics, Uzbekistan Academy of Sciences
  9, University str., 100174, Tashkent, Uzbekistan;
 \item[] $^b$ National University of Uzbekistan, 4, University str., 
  Tashkent, 100174, Uzbekistan.
  \end{itemize}}
\email{haydarov\_imc@mail.ru}
	
	\address{%
  B.A. Omirov$^{c,d}$ 
   \begin{itemize}
 \item[]  $^c$ Suzhou Research Institute, Harbin Institute of Technology, Suzhou 215104, China;
\item[] $^d$ Institute for Advanced Study in Mathematics, Harbin Institute of Technology, Harbin 150001, China.
\end{itemize}}
\email{omirovb@mail.ru}
	\address{%
  U.A. Rozikov$^{a,b,d}$ 
   \begin{itemize}
 \item[] $^a$ V.I. Romanovskiy Institute of Mathematics, Uzbekistan Academy of Sciences
  9, Universitet str., 100174, Tashkent, Uzbekistan;
 \item[]  $^b$ National University of Uzbekistan, 4, University str. Tashkent, 100174, Uzbekistan;
  \item[]  $^d$ Institute for Advanced Study in Mathematics, Harbin Institute of Technology, Harbin 150001, China.
  \end{itemize}}
\email{rozikovu@yandex.ru}
	
	\begin{abstract} We introduce a {\it non-linear} generalization of the classical Dobrushin-Lanford-Ruelle (DLR) framework 
by developing the concept of a $q$-specification and the associated $q$-equilibrium measures. These objects arise naturally from a family of non-linear $q$-stochastic operators acting on the space of probability measures. A $q$-equilibrium measure is characterized as a fixed point of such operators, providing a non-linear analogue of the Gibbs equilibrium in the sense of DLR. We establish general conditions ensuring the existence and uniqueness of $q$-equilibrium measures and demonstrate how quasilocality plays a decisive role in their construction. Moreover, we exhibit examples of $q$-specifications with an empty set of $q$-equilibrium measures. We characterize the set of $q$-equilibrium measures by studying the dynamical systems generated by a class of 
$q$-stochastic operators. As a concrete application, we show that for the one-dimensional Ising model at sufficiently low temperatures, multiple $q$-equilibrium measures may exist, even though the classical Gibbs measure remains unique. Our results reveal that the $q$-specification formalism extends the DLR theory from linear to non-linear settings and opens a new direction in the study of Gibbs measures and equilibrium states of physical systems.
	\end{abstract}
	\maketitle
{\bf Mathematics Subject Classifications (2020).} 82B20, 60J20, 37A60, 47H10.

{\bf{Key words.}} $q$-stochastic operator; $q$-specification; $q$-equilibrium measure; non-linear DLR equation; Gibbs measure; quasilocality;  Ising model.

\tableofcontents
\section{Introduction}

The DLR equations form one of the cornerstones of modern statistical mechanics. 
They provide a rigorous characterization of equilibrium states for lattice systems with local interactions through Gibbs measures, 
unifying probabilistic, analytic, and thermodynamic viewpoints. 
Within this framework, a Gibbs measure is  a probability measure on the configuration space whose conditional distributions satisfy a system of \emph{linear} consistency equations determined by a local specification. 
Since the pioneering works in the late 1960s, the DLR formalism has become a standard language in the mathematical theory of phase transitions, interacting particle systems, and ergodic theory (see, e.g., \cite{Enter, FV, Ge, Rue}).

Beyond its classical role in equilibrium statistical mechanics, the DLR approach has influenced a broad range of mathematical structures, 
from Markov random fields and stochastic partial differential equations to probabilistic graphical models in information theory and machine learning \cite{AR}, \cite{Beh}, \cite{HRV}, \cite{Krza}. 

In the classical setting, the DLR operator associated with a given specification acts linearly on the space of probability measures. 
This linearity reflects the additivity of physical interactions and ensures that convex combinations of Gibbs measures are again Gibbs measures -- a property that underlies much of the classical theory of phase transitions and equilibrium statistical mechanics \cite{Do,  Ge, Rue}. Similarly, in Markov processes \cite{SK}, stochastic differential equations, statistical physics (e.g., transfer operators), and machine learning (e.g., probabilistic graphical models), a stochastic operator is typically a linear operator mapping the set of probability measures to itself. 

However, numerous natural systems in physics, biology, and information sciences exhibit non-linear aggregation laws, in which the measure describing the system's state satisfies a non-linear equation. Evolution of such systems described by non-linear stochastic operators (see \cite{Bar,  MG, MSQ, Rab, Rpd, RS}, and references therein). They often induce non-trivial dynamical systems on the space of measures, whose fixed points correspond to equilibrium distributions of complex, self-interacting systems, 
providing a natural generalization of classical Gibbs measures.
Examples include multi-locus genetic models \cite{Ev, Ly, MG, OR, RRS} and non-linear classical stochastic processes \cite{Bo, Fr1, Gr, Reh}, where interactions cannot be captured by linear superposition. 
In such systems, the classical linear DLR equations no longer fully describe equilibrium behavior, motivating the development of non-linear extensions of the Gibbs formalism.

It is worth mentioning that nonlinear extensions of the Gibbs measure formalism were already considered in recent paper \cite{CP}, where a natural class of nonlinear dynamics for spin systems, such as the Ising model, was introduced. These dynamics were motivated by the importance of studying the evolution of systems of interacting entities under pairwise interactions and capture a number of significant nonlinear models from various fields, including chemical reaction networks, Boltzmann's model of an ideal gas, recombination in population genetics, and genetic algorithms (see \cite{CP2}, \cite{CS}, \cite{CP} and references therein).

This perspective suggests that many systems of current interest -- from evolutionary dynamics to non-linear 
inference in machine learning -- may be fruitfully analyzed using a \textbf{non-linear DLR}-type framework, 
which can reveal multiple or novel equilibrium states not captured by the linear theory.

The main problems addressed in this paper concern the extension of the DLR framework to a non-linear setting. 
We introduce a family of $q$-stochastic operators and define the corresponding $q$-specifications, which generalize the classical linear specifications of DLR theory. 
This framework retains the essential probabilistic and variational features of Gibbs measures while incorporating non-linear dependence on states of local configurations. 
In particular, for $q=1$ we recover the classical DLR equations, whereas for $q\ge 2$ we obtain genuinely new types of non-linear consistency relations.

The motivation for studying such non-linear generalizations is twofold. 
From a theoretical standpoint, they reveal a new hierarchy of 
equilibrium structures that extend the traditional Gibbsian formalism to non-linear operator settings, thereby connecting statistical mechanics with non-linear probability theory. 
From an applied viewpoint, these equations may describe equilibrium states of systems with non-additive entropy, mean-field feedback, or multi-agent interactions, 
relevant in phase transitions, evolutionary dynamics, and probabilistic inference models. 
Thus, the proposed $q$-specification framework opens a new direction in the theory of Gibbs measures and equilibrium states of physical systems.\\

The paper is organized as follows.
In {\it Section 2}, we recall definitions and results related to $q$-stochastic operators and Gibbs measures.
{\it Section 3}, introduces the concept of a $q$-equilibrium measure via a new notion of $q$-specification, which is associated with a family of  non-linear $q$-stochastic operators. It is noteworthy that a $q$-equilibrium measure coincides with a fixed point of each of these non-linear operators. Furthermore, we construct a Gibbs $q$-specification that reduces to the classical Gibbs specification when $q=1$. For $q \ge 2$, however, our construction extends the traditional DLR linear equations to a {\bf non-linear} framework.
In {\it Section 4}, we define the notion of quasilocality for $q$-specifications and show that this property guarantees the non-emptiness of the corresponding set of $q$-equilibrium measures.
{\it Section 5} is devoted to the problem of uniqueness of such measures, 
analyzed through their sensitivity to boundary conditions.
In {\it Section 6}, we provide an example of a $q$-specification for which the set of $q$-equilibrium measures is empty.
{\it Section 7} deals with the dynamical systems generated by $q$-stochastic operators and their associated $q$-equilibrium measures; in this section we establish a characterization theorem describing the entire set of $q$-equilibrium measures corresponding to a given $q$-specification.
Finally, in {\it Section 8}, we apply our general theory to the one-dimensional Ising model and show that, for sufficiently low temperatures, multiple $q$-equilibrium measures exist. Since for this model the ordinary Gibbs measure is unique (that is, the linear DLR equations admit a unique solution), this result demonstrates that our non-linear DLR-type equations may admit more than one solution.

\section{Preliminaries}
	\subsection{Basic Setup} 
	
	Let $V$ be a countable set, which could be the vertex set of a graph, e.g., $V = \mathbb{Z}^d$ with $d \in \mathbb{N}$. 
    Consider  a Polish space $\Omega_0$ with sigma-algebra $\mathcal{F}_0$. We call $\Omega_0$ the local state space; in this paper we consider a finite set $\Omega_0$.
	
	 For any sub-volume $\Lambda \subset V$ (possibly infinite) define
	$$
	\Omega_\Lambda = \Omega_0^\Lambda = \{ (\sigma_x)_{x \in \Lambda} : \sigma_x \in \Omega_0 \}.
	$$
	When $\Lambda = V$, write $\Omega = \Omega_V$. The measurable structure is given by the product sigma-algebra
	$$
	\mathcal{B}_\Lambda = \bigotimes_{x \in \Lambda} \mathcal{F}_0.
	$$
	
	For $x \in V$, the projection onto $x$ is $\gamma_x : \Omega \to \Omega_0$. The restriction of $\sigma \in \Omega$ to $\Lambda \subset V$ is $\sigma_\Lambda \in \Omega_\Lambda$. The concatenation $\xi\eta \in \Omega$ for $\xi \in \Omega_\Lambda$, $\eta \in \Omega_{\Lambda^c}$ satisfies $\sigma_\Lambda(\xi\eta) = \xi$ and $\sigma_{\Lambda^c}(\xi\eta) = \eta$.
	
	 Write $\Lambda \Subset V$ if $\Lambda$ is finite. For  $\Lambda \Subset V$, the sigma-algebra of cylinders is
	$\mathcal{C}_{\Lambda} := \gamma_\Lambda^{-1}(\mathcal{B}_\Lambda)$. For $\Delta \subset V$, the sigma-algebra $\mathcal{F}_\Delta$ is $\mathcal{F}_\Delta := \sigma(\mathcal{C}_\Delta).$ When $\Delta = V$, write $\mathcal{F} = \mathcal{F}_V = \bigotimes_{i \in V} \mathcal{F}_0$.
	
	The $q$-th power of $\mathcal F$ is the product $\sigma$-algebra
	$$
	\mathcal F^q := \underbrace{\mathcal F \otimes \mathcal F \otimes \cdots \otimes \mathcal F}_{q \ \text{times}},
	$$
	which is the smallest $\sigma$-algebra on $\Omega^q$ containing all measurable sets
	$$
	A_1 \times A_2 \times \cdots \times A_q, \quad A_i \in \mathcal F.
	$$
\subsection{Stochastic operators} 

 In the theory of Markov chains, the transition matrix $P$ defines 
 a stochastic operator that advances the distribution of states over time.

Let $L^1$ be  space of integrable functions, 
a stochastic operator $T: L^1 \to L^1$ is linear ($T(af + bg) = aTf + bTg$);
positive ($f \geq 0 \Rightarrow Tf \geq 0$), norm-preserving ($\|Tf\|_1 = \|f\|_1$, which ensures total probability is preserved).

On an infinite-dimensional space a stochastic operator can be given as
$$
(Tf)(x) = \int K(x, y) f(y) \, dy,
$$
where $K(x, y) \geq 0$ and $\int K(x, y) \, dx = 1$ for each $y$.

For example, non-linear Markov operators generalize classical 
Markov operators, that is instead of
$$
T(\mu)(A) = \int P(x, A) \, d\mu(x)
$$
one may have
$$
T(\mu)(A) = \int P(x, \mu, A) \, d\mu(x)
$$
where $P(x, \mu, A)$ depends non-linearly on $\mu$.

In this paper, we examine a specific class of nonlinear stochastic 
operators called as $q$-stochastic.

\subsection{$q$-Stochastic operators and their connections to known transformations}

Denote by $\mathcal{M}_1(\Omega, {\mathcal F})$ the set of all probability measures on a measurable space $(\Omega, {\mathcal F})$. 

\begin{defn}\label{gd1} Let $q\in \mathbb N$. A mapping $V_q : \mathcal M_1(\Omega,{ \mathcal F})\to \mathcal M_1(\Omega, { \mathcal F})$ is called a $q$-stochastic operator ($q$-SO) if, for an arbitrary measure $\l\in \mathcal M_1(\Omega,{ \mathcal F}),$ the measure $\l'=V_q(\l)$ is defined as 
\begin{equation}\label{Vq} V_q(\l): \ \ \l'(A)=\int_{\Omega^q} P(x_1,x_2,\dots, x_q, A)\prod_{i=1}^q\l(dx_i), \ \ \forall A\in { \mathcal F}, \end{equation} 
where $P(x_1, \dots, x_q, A)$ satisfies the following conditions: 
\begin{itemize}
\item[(i)] $P(x_1, \dots, x_q,\cdot)\in \mathcal M_1(\Omega,{ \mathcal F})$ for any fixed $x_1, \dots, x_q \in \Omega;$ 

\item[(ii)] $P(x_1, \dots, x_q, A),$ regarded as a function of $q$ variables $x_1$, $\dots$ $x_q,$ is measurable on $(\Omega^q, { \mathcal F^{\otimes q}})$ for any fixed $A\in { \mathcal F}.$ 
\end{itemize}
\end{defn}

For finite set $\Omega$, say $\Omega = \{1, 2, \dots, n\}$, the space $\mathcal{M}_1(\Omega, {\mathcal F})$ can be identified with the simplex 
$$
S^{n-1} = \Bigl\{ x = (x_1, \dots, x_n) \in \mathbb{R}^n \ \big|\ x_i \ge 0, \ \sum_{i=1}^n x_i = 1 \Bigr\}.
$$
In this case, the operator $V_q$ acts as
\begin{equation}\label{eq:discreteVq}
x_k' = \sum_{i_1, \dots, i_q = 1}^n P_{i_1 \dots i_q, k}\, x_{i_1} \dots x_{i_q},
\end{equation}
where $P_{i_1 \dots i_q, k} = P(i_1, \dots, i_q, \{k\})$ and
\begin{equation}\label{eq:coeff}
P_{i_1 \dots i_q, k} \ge 0, 
\qquad 
\sum_{k=1}^n P_{i_1 \dots i_q, k} = 1, 
\qquad 
\forall i_1, \dots, i_q \in \Omega.
\end{equation}
The mapping $V_q$ describes the evolution of probability measures in systems with interactions among $q$ components. It is non-linear for all $q \ge 2$ and naturally arises in statistical mechanics, population genetics, and dynamical systems as a model of nonlinear stochastic evolution~\cite{Bernstein1924, Ly, Rpd}. 

The $q$-SO offers a unified framework for studying non-linear transformations of probability measures. It relates to several classical concepts: renormalization group transformations act on Gibbs measures by coarse-graining ~\cite{Enter};  Glauber and Kawasaki dynamics describe stochastic evolutions of Gibbs measures that may become non-Gibbs after a quench~\cite{Liggett1985, KipnisLandim1999}; in disordered systems, random interactions often destroy Gibbsian structure~\cite{Bovier2006}; Decimation and coarse-graining ~\cite{GriffithsPearce1978, BleherSinai1976}. Analogously, the $q$-SO integrates over $q$-tuples of sites, 
potentially inducing Gibbs-non-Gibbs transitions ~\cite{Fe}.

\subsection{Gibbs Measures}
In order to define Gibbs measures, we need some additional notations  \cite{K}, \cite{She}.

Let $(X, \mathcal X)$ and $(Y, \mathcal Y)$ be general measure spaces.

A function $\pi : \mathcal X\times Y\to [0, \infty]$ is called a probability kernel from $(Y, \mathcal Y)$ to $(X, \mathcal X)$ if
\begin{itemize}
	\item[1.] $\pi(\cdot | y)$ is a probability measure on $(X, \mathcal X)$ for each fixed $y\in Y$;
	\item[2.] $\pi(A | \cdot)$ is $\mathcal Y$-measurable for each fixed $A\in \mathcal X$.
\end{itemize}

Such a kernel maps each measure $\mu$, on $(Y, \mathcal Y)$ to a measure $\mu\pi$ on $(X, \mathcal X)$ by
$$\mu\pi(A)=\int \pi(A | \cdot)d\mu.$$

Note that for the case $q=1$ the function $P(x_1, \dots, x_q,\cdot)$ is a probability kernel. 

Consider a $G=(V,L)$, where $V$ is a countable set of vertices and $L$ is the set of edges. Take a finite set $\Psi=\{1,2,...,q\}$, $q\geq 2$. This is the set of spin values.
For $A\subseteq V$ a spin {\it configuration} $\sigma_A$
on $A$ is defined as a function $$x\in
A\to\sigma_A(x)\in\Psi.$$

The set of all configurations given on $A$ is denoted by $\Omega_A$.
We denote $\Omega=\Omega_V$ and
$\sigma=\sigma_V.$

The cylinder $\sigma$-algebra  is the one that is
generated by cylinder sets.

The $\sigma$-algebra generated by cylinders with base contained in $A$ is denoted by 
$\mathcal F_A$, this consists of all the events that depend only on the spins inside $A$.

Consider a family $\Phi=\{\Phi_\Lambda: \Lambda\subset V \}$ of measurable potential 
functions $\Phi_\Lambda:\Omega\to \mathbb R\cup\{\infty\}$
(one for each finite subset $\Lambda$ of $V$)\ each $\Phi_\Lambda$ is $\mathcal F_\Lambda$
measurable. 
We say $\sigma$ has finite energy if $ \Phi_\Lambda(\sigma) < \infty$ for all finite $\Lambda$.

The following is a probability kernel from $(\Omega, \mathcal F_{V\setminus\Lambda})$ to $(\Omega, \mathcal F)$:
$$\gamma_\Lambda(A,\omega)=Z_\Lambda(\omega)^{-1}\int\exp(-H_\Lambda(\sigma_\Lambda\omega_{\Lambda^c}))
\mathbf{1}_A(\sigma_\Lambda\omega_{\Lambda^c})\nu^{\otimes\Lambda}(d\sigma_\Lambda),$$
where $\nu=\{\nu(i)>0, i\in E\}$ is a counting measure,
$$\nu^{\otimes\Lambda}(d\sigma_\Lambda)=\prod_{x\in \Lambda}\nu(d\sigma_\Lambda(x)) \ \ \mbox{and}$$
$$Z_\Lambda(\omega)=\int\exp(-H_\Lambda(\sigma_\Lambda\omega_{\Lambda^c}))\nu^{\otimes\Lambda}(d\sigma_\Lambda).$$
 We say $\sigma$ is $\Phi$-admissible
if each $Z_\Lambda(\sigma)$ is finite and non-zero.

Given a measure $\mu$ on $(\Omega, \mathcal F)$, we define a new
measure $\mu\gamma_\Lambda$ by
$$\mu\gamma_\Lambda(A)=\int\gamma_\Lambda(A, \cdot)d\mu$$

\begin{defn} A probability measure $\mu$ on $(\Omega, \mathcal F)$ is called a Gibbs measure if $\mu$ is supported on the set of $\Phi$-admissible configurations in $\Omega$ and for all finite subset $\Lambda$ we have
	$\mu \gamma_\Lambda=\mu.$
\end{defn}

A Gibbs measure used to describe the states of a system in statistical mechanics, characterized by local interactions defined by a Hamiltonian. The above definition says that a Gibbs measure satisfies a local specification property known as the DLR condition, meaning that the conditional probability of a spin configuration, given the states of its neighbors, follows a specific form dictated by the Hamiltonian.

Denote by $\mathcal G(\gamma)$ the set of all Gibbs measures corresponding to $\gamma=\{\gamma_\Lambda: \Lambda\subset V\}$. This $\gamma$ is called a Gibbs specification.

\section{$q$-equilibrium measures}

\subsection{Probability $q$-Kernels and $q$-Specifications}

For a probability measure $\mu$ on $(\Omega,\mathcal F)$ and a bounded
measurable function $f:\Omega^q\to\mathbb R$, we write
\begin{equation}\label{fm}
	\mu^{\otimes q}(f)
	=
	\int_{\Omega^q}
	f(\omega_1,\ldots,\omega_q)
	\prod_{i=1}^q \mu(d\omega_i).
\end{equation}

\begin{rk}\label{xf}
	If $f=\mathbf 1_A$ is the indicator of a set
	\[
	A=B^q
	=
	\underbrace{B\times B\times\cdots\times B}_{q\text{ times}}
	\subseteq\Omega^q,
	\qquad B\in\mathcal F,
	\]
	then
	\[
	\mu^{\otimes q}(f)
	=
	\mu^{\otimes q}(A)
	=
	\mu^{\otimes q}(B^q)
	=
	\prod_{i=1}^q\mu(B)
	=
	\bigl(\mu(B)\bigr)^q.
	\]
\end{rk}

\begin{defn}
	Let $\Lambda\subset V$. A \textit{probability $q$-kernel} from
	$\mathcal F_{\Lambda^c}^q$ to $\mathcal F$ is a map $
	\gamma_\Lambda:\mathcal F\times\Omega^q\longrightarrow[0,1]$
	such that:
	\begin{enumerate}
		\item[(i)] for every
		$(\eta_1,\ldots,\eta_q)\in\Omega^q$, the map
		$
		\gamma_\Lambda(\,\cdot\mid\eta_1,\ldots,\eta_q)
		$
		is a probability measure on $(\Omega,\mathcal F)$;
		
		\item[(ii)] for every $A\in\mathcal F$, the map $
		(\eta_1,\ldots,\eta_q)
		\longmapsto
		\gamma_\Lambda(A\mid\eta_1,\ldots,\eta_q)
		$
		is $\mathcal F_{\Lambda^c}^q$-measurable.
	\end{enumerate}
	
	The kernel is called \textit{proper} if
	\begin{equation}\label{proper}
		\gamma_\Lambda(A\mid\eta_1,\ldots,\eta_q)
		=
		\frac{1}{q}\sum_{i=1}^q
		\mathbf 1_A(\eta_i),
		\qquad
		A\in\mathcal F_{\Lambda^c}.
	\end{equation}
	Thus, on events measurable with respect to the outside
	$\Lambda^c$, the kernel agrees with the empirical average of the
	boundary indicators.
\end{defn}

For $\mu\in\mathcal M_1(\Omega,\mathcal F)$ and a probability
$q$-kernel $\gamma_\Lambda$, define the probability measure
$\mu^{\otimes q}\gamma_\Lambda$ on $(\Omega,\mathcal F)$ by
\begin{equation}\label{mugamma}
	(\mu^{\otimes q}\gamma_\Lambda)(A)
	=
	\int_{\Omega^q}
	\gamma_\Lambda(A\mid\eta_1,\ldots,\eta_q)
	\prod_{i=1}^q\mu(d\eta_i),
	\qquad
	A\in\mathcal F.
\end{equation}

For a bounded $\mathcal F^q$-measurable function
$f:\Omega^q\to\mathbb R$, define
\begin{equation}\label{fl}
	\gamma_\Lambda^{\otimes q}f
	(\eta_1,\ldots,\eta_q)
	=
	\int_{\Omega^q}
	f(\sigma_1,\ldots,\sigma_q)
	\prod_{i=1}^q
	\gamma_\Lambda(d\sigma_i\mid\eta_1,\ldots,\eta_q).
\end{equation}

For two probability $q$-kernels $\gamma_\Delta$ and $\gamma_\Lambda$,
their composition is defined by
\begin{equation}\label{gg}
	\begin{split}
		(\gamma_\Delta\gamma_\Lambda)
		(A\mid\eta_1,\ldots,\eta_q)
		&:=
		\gamma_\Delta^{\otimes q}
		\bigl(
		\gamma_\Lambda(A\mid\cdot)
		\bigr)
		(\eta_1,\ldots,\eta_q)
		\\
		&=
		\int_{\Omega^q}
		\gamma_\Lambda(A\mid\zeta_1,\ldots,\zeta_q)
		\prod_{i=1}^q
		\gamma_\Delta(d\zeta_i\mid\eta_1,\ldots,\eta_q),
	\end{split}
\end{equation}
whenever the composition is defined.

Notice that, for $q=1$, \eqref{gg} reduces to the usual composition of
probability kernels:
\[
(\gamma_\Delta\gamma_\Lambda)(A\mid\eta)
=
\int_\Omega
\gamma_\Lambda(A\mid\zeta)
\gamma_\Delta(d\zeta\mid\eta).
\]

\begin{rk}\label{ggg}
	For $\Delta=\Lambda$, the composition \eqref{gg} becomes
	\begin{equation}\label{eq:gamma-square}
		(\gamma_\Lambda\gamma_\Lambda)
		(A\mid\eta_1,\ldots,\eta_q)
		=
		\int_{\Omega^q}
		\gamma_\Lambda(A\mid\zeta_1,\ldots,\zeta_q)
		\prod_{i=1}^q
		\gamma_\Lambda(d\zeta_i\mid\eta_1,\ldots,\eta_q).
	\end{equation}
	
	For $q>1$, a probability $q$-kernel is not necessarily idempotent under
	the composition \eqref{gg}; that is,
	\begin{equation}\label{eq:not-idempotent}
		\gamma_\Lambda\gamma_\Lambda
		\neq
		\gamma_\Lambda
	\end{equation}
	may occur.
	
	Indeed, consider $q=2$ and $\Omega=\{0,1\}$, $\mathcal F=2^\Omega.$
	Define a probability $2$-kernel by
	\begin{equation}\label{eq:counterexample-kernel}
		\gamma_\Lambda(\{1\}\mid\eta_1,\eta_2)
		=
		\begin{cases}
			1,&(\eta_1,\eta_2)=(0,0),\\[1mm]
			0,&\text{otherwise}.
		\end{cases}
	\end{equation}
	Equivalently,
	\[
	\gamma_\Lambda(\,\cdot\mid0,0)=\delta_1,
	\qquad
	\gamma_\Lambda(\,\cdot\mid1,1)=\delta_0.
	\]
	Consequently,
	\[
	\gamma_\Lambda^{\otimes2}
	(\,\cdot\mid0,0)
	=
	\delta_1\otimes\delta_1
	=
	\delta_{(1,1)}.
	\]
	Hence, by \eqref{eq:gamma-square},
	\[
			(\gamma_\Lambda\gamma_\Lambda)
		(\{1\}\mid0,0)
		=
		\int_{\Omega^2}
		\gamma_\Lambda(\{1\}\mid\zeta_1,\zeta_2)
		\,\delta_{(1,1)}(d\zeta_1,d\zeta_2)
		=
		\gamma_\Lambda(\{1\}\mid1,1)=
		0.
		\]
	On the other hand, $
	\gamma_\Lambda(\{1\}\mid0,0)=1.
	$
	Therefore,
	\[
	(\gamma_\Lambda\gamma_\Lambda)(\{1\}\mid0,0)
	\neq
	\gamma_\Lambda(\{1\}\mid0,0),
	\]
	which proves \eqref{eq:not-idempotent}.
	
	Thus, unlike the ordinary case $q=1$, idempotence does not follow from
	the definition of a probability $q$-kernel alone.
\end{rk}

For $\eta\in\Omega$, denote
\[
\Omega_\Lambda^\eta
=
\{\sigma\in\Omega:
\sigma_{\Lambda^c}=\eta_{\Lambda^c}\},
\]
and, for
$(\eta_1,\ldots,\eta_q)\in\Omega^q$, define
\[
\Omega_\Lambda^{\eta_1,\ldots,\eta_q}
=
\bigcup_{i=1}^q
\Omega_\Lambda^{\eta_i}.
\]

\begin{lemma}\label{sup}
	Let $\gamma_\Lambda$ be a proper probability $q$-kernel. Then, for
	every $(\eta_1,\ldots,\eta_q)\in\Omega^q$, the measure $
	\gamma_\Lambda(\,\cdot\mid\eta_1,\ldots,\eta_q)$
	is supported on
	$\Omega_\Lambda^{\eta_1,\ldots,\eta_q}$.
\end{lemma}

\begin{proof}
	The set
	$\Omega_\Lambda^{\eta_1,\ldots,\eta_q}$ belongs to
	$\mathcal F_{\Lambda^c}$. Moreover,
	\[
	\eta_i\in
	\Omega_\Lambda^{\eta_1,\ldots,\eta_q},
	\qquad i=1,\ldots,q.
	\]
	Therefore, by properness,
	\[
			\gamma_\Lambda
		\left(
		\Omega_\Lambda^{\eta_1,\ldots,\eta_q}
		\mid
		\eta_1,\ldots,\eta_q
		\right)=
		\frac1q
		\sum_{i=1}^q
		\mathbf 1_{\Omega_\Lambda^{\eta_1,\ldots,\eta_q}}(\eta_i)
		=
		\frac1q\sum_{i=1}^q1
		=
		1.
		\]
	Thus the kernel assigns full mass to
	$\Omega_\Lambda^{\eta_1,\ldots,\eta_q}$.
\end{proof}

\begin{defn}
	A \textit{$q$-specification} is a family $
	\gamma=(\gamma_\Lambda)_{\Lambda\Subset V}$
	of proper probability $q$-kernels satisfying the consistency condition
	\begin{equation}\label{q-consistency}
		\gamma_\Delta\gamma_\Lambda
		=
		\gamma_\Delta,
		\qquad
		\Lambda\subseteq\Delta\Subset V.
	\end{equation}
	
	In particular, taking $\Delta=\Lambda$ in \eqref{q-consistency}, every
	kernel belonging to a $q$-specification is idempotent:
	\begin{equation}\label{idempotence}
		\gamma_\Lambda\gamma_\Lambda
		=
		\gamma_\Lambda,
		\qquad
		\Lambda\Subset V.
	\end{equation}
	
	A probability measure $\mu\in\mathcal M_1(\Omega,\mathcal F)$ is called
	\textit{compatible} with $\gamma$ if
	\begin{equation}\label{DLR}
		\mu^{\otimes q}\gamma_\Lambda
		=
		\mu,
		\qquad
		\Lambda\Subset V.
	\end{equation}
	Such a measure is called a \textit{$q$-equilibrium measure}. The set of
	all $q$-equilibrium measures is denoted by $\mathcal G_q(\gamma)$.
	\end{defn}

The following lemma gives a useful consequence of properness.

\begin{lemma}\label{mg}
	Let $\gamma_\Lambda$ be a proper probability $q$-kernel. Fix
	$A\in\mathcal F$ and $B\in\mathcal F_{\Lambda^c}$. For
	$(\omega_1,\ldots,\omega_q)\in\Omega^q$, let
	\[
	\nu(\,\cdot\,)
	=
	\gamma_\Lambda(\,\cdot\mid\omega_1,\ldots,\omega_q)
	\]
	and define
	\[
	p_B(\omega_1,\ldots,\omega_q)
	=
	\gamma_\Lambda(B\mid\omega_1,\ldots,\omega_q)
	=
	\frac1q\sum_{i=1}^q\mathbf 1_B(\omega_i).
	\]
	Then:
	\begin{enumerate}
		\item[(1)]
		$
		0\le
		\nu(A\cap B)
		\le
		p_B(\omega_1,\ldots,\omega_q),
		\qquad
		\nu(A\cap B)\le\nu(A);
		$
				\item[(2)] if
		$
		p_B(\omega_1,\ldots,\omega_q)=0,
		$
		then
		$
		\nu(A\cap B)=0;
		$
				\item[(3)] if
		$
		p_B(\omega_1,\ldots,\omega_q)=1,
		$
		then
		$
		\nu(A\cap B)=\nu(A).
		$
	\end{enumerate}
\end{lemma}
\begin{proof}
	Since $	A\cap B\subseteq B$ and $A\cap B\subseteq A$,
	we have
	\[
	0\le\nu(A\cap B)\le\nu(B)
	=
	p_B(\omega_1,\ldots,\omega_q)
	\]
	and
	$
	\nu(A\cap B)\le\nu(A).
	$
	This proves (1).
	
	If
	$p_B(\omega_1,\ldots,\omega_q)=0$, then
	$
	\nu(B)=0.
	$
	Since $A\cap B\subseteq B$,
	$
	0\le\nu(A\cap B)\le\nu(B)=0,
	$
	and therefore
	$
	\nu(A\cap B)=0.
	$
	This proves (2).
	
	If
	$p_B(\omega_1,\ldots,\omega_q)=1$, then
	$
	\nu(B)=1$ and hence $\nu(B^c)=0$.
	Since
	$$
	A=(A\cap B)\cup(A\cap B^c),
	$$
	we obtain
	$$
			\nu(A)
		=
		\nu(A\cap B)+\nu(A\cap B^c)
		\le
		\nu(A\cap B)+\nu(B^c)
		=
		\nu(A\cap B).
		$$
	Since $A\cap B\subseteq A$, we also have
	$
	\nu(A\cap B)\le\nu(A).
	$
	Thus
	$
	\nu(A\cap B)=\nu(A).
	$
\end{proof}

\begin{rk}
	For each $\Lambda\Subset V$, define the associated
	$q$-stochastic operator
	\[
	V_{q,\Lambda}:
	\mathcal M_1(\Omega,\mathcal F)
	\longrightarrow
	\mathcal M_1(\Omega,\mathcal F)
	\]
	by
	\begin{equation}\label{qSO}
		V_{q,\Lambda}(\mu)(A)
		=
		\int_{\Omega^q}
		\gamma_\Lambda(A\mid\omega_1,\ldots,\omega_q)
		\prod_{i=1}^q\mu(d\omega_i),
		\qquad
		A\in\mathcal F.
	\end{equation}
	Then
	\[
	V_{q,\Lambda}(\mu)
	=
	\mu^{\otimes q}\gamma_\Lambda,
	\]
	and the compatibility condition \eqref{DLR} is equivalent to
	\[
	V_{q,\Lambda}(\mu)=\mu,
	\qquad
	\Lambda\Subset V.
	\]
	Thus a $q$-equilibrium measure is a common fixed point of the family
	of nonlinear operators
	$\{V_{q,\Lambda}\}_{\Lambda\Subset V}$.
	
	This provides a nonlinear analogue of the DLR characterization of
	Gibbs measures. In the special case $q=1$, we have
	\[
	V_{1,\Lambda}(\mu)(A)
	=
	\int_\Omega
	\gamma_\Lambda(A\mid\omega)\,\mu(d\omega)
	=
	(\mu\gamma_\Lambda)(A),
	\]
	and the compatibility condition becomes the usual DLR equation
	\[
	\mu\gamma_\Lambda=\mu.
	\]
	Consequently, if
	$(\gamma_\Lambda)_{\Lambda\Subset V}$ is an ordinary Gibbs
	specification, then the $1$-equilibrium measures coincide with the
	usual Gibbs measures.
	
	The $q$-equilibrium framework therefore provides a nonlinear extension
	of the classical Gibbsian formalism, in which the local transformation
	depends on $q$ independently sampled configurations rather than on a
	single configuration.
\end{rk}

\subsection{Gibbs $q$-specification}

In the context of $q$-stochastic operators, we introduce a Gibbs-type
probability $q$-kernel based on a finite-volume Hamiltonian. Let
$V$ be a countable set, let $\Omega_0$ be a measurable spin space
equipped with a reference measure $\nu$, and let
$\Omega=\Omega_0^V$. For a finite set $\Lambda\Subset V$, write
$\Omega_\Lambda=\Omega_0^\Lambda$ and let
$\nu^{\otimes\Lambda}$ denote the corresponding product reference
measure.

For $(\omega_1,\ldots,\omega_q)\in\Omega^q$, define
\begin{equation}\label{ke}
	\begin{aligned}
		\gamma_{q,\Lambda}(A\mid\omega_1,\ldots,\omega_q)
		&=
		\frac{1}{Z_\Lambda(\omega_1,\ldots,\omega_q)}
		\int_{\Omega_\Lambda}
		\exp\Big(
		-H_\Lambda\big(
		\sigma_\Lambda;
		\omega_{1,\Lambda^c},\ldots,\omega_{q,\Lambda^c}
		\big)
		\Big)
		\\
		&\qquad\qquad\times
		\left(
		\frac{1}{q}\sum_{i=1}^q
		\mathbf{1}_A(\sigma_\Lambda\omega_{i,\Lambda^c})
		\right)
		\nu^{\otimes\Lambda}(d\sigma_\Lambda),
	\end{aligned}
\end{equation}
where $A\in\mathcal F$, and
\begin{equation}\label{Zq}
	Z_\Lambda(\omega_1,\ldots,\omega_q)
	=
	\int_{\Omega_\Lambda}
	\exp\Big(
	-H_\Lambda\big(
	\sigma_\Lambda;
	\omega_{1,\Lambda^c},\ldots,\omega_{q,\Lambda^c}
	\big)
	\Big)
	\nu^{\otimes\Lambda}(d\sigma_\Lambda)
\end{equation}
is the corresponding partition function. Here
$H_\Lambda$ is a finite-volume Hamiltonian which may depend on the
$q$ boundary configurations through their restrictions to
$\Lambda^c$. The assumptions below ensure that
\eqref{ke} defines a probability $q$-kernel.

\begin{pro}
	Suppose that, for every finite $\Lambda\Subset V$,
	
	\begin{enumerate}
		\item the function
		\[
		(\sigma_\Lambda,\omega_{1,\Lambda^c},\ldots,\omega_{q,\Lambda^c})
		\longmapsto
		H_\Lambda(
		\sigma_\Lambda;
		\omega_{1,\Lambda^c},\ldots,\omega_{q,\Lambda^c})
		\]
		is measurable;
		
		\item for every
		$(\omega_1,\ldots,\omega_q)\in\Omega^q$,
		\[
		0<
		Z_\Lambda(\omega_1,\ldots,\omega_q)
		<\infty.
		\]
	\end{enumerate}
	
	Then \eqref{ke} defines a probability $q$-kernel
	\[
	\gamma_{q,\Lambda}:
	\mathcal F\times\Omega^q\longrightarrow[0,1],
	\]
	which is proper with respect to $\mathcal F_{\Lambda^c}$. Moreover, if
	the family
	$\gamma=(\gamma_{q,\Lambda})_{\Lambda\Subset V}$ satisfies the
	consistency condition
	\begin{equation}\label{qconsG}
		\gamma_{q,\Delta}\gamma_{q,\Lambda}
		=
		\gamma_{q,\Delta},
		\qquad
		\Lambda\subseteq\Delta\Subset V,
	\end{equation}
	under the composition defined above, then $\gamma$ is a
	$q$-specification. In this case, we call $\gamma$ a Gibbs
	$q$-specification.
\end{pro}

\begin{proof}
	Fix $\Lambda\Subset V$ and
	$(\omega_1,\ldots,\omega_q)\in\Omega^q$.
	
	First, by the measurability assumption on $H_\Lambda$, the integrand
	in \eqref{ke} is measurable as a function of $\sigma_\Lambda$.
	Assumption (2) implies that the normalization constant
	$Z_\Lambda(\omega_1,\ldots,\omega_q)$ is strictly positive and finite.
	Hence \eqref{ke} is well-defined.
	
	For $A=\Omega$, we have
	\[
	\frac{1}{q}\sum_{i=1}^q
	\mathbf{1}_\Omega(\sigma_\Lambda\omega_{i,\Lambda^c})
	=1,
	\]
	and therefore
	\[
			\gamma_{q,\Lambda}(\Omega\mid\omega_1,\ldots,\omega_q)=
		\frac{1}{Z_\Lambda(\omega_1,\ldots,\omega_q)}
		\int_{\Omega_\Lambda}
		e^{-H_\Lambda(
			\sigma_\Lambda;
			\omega_{1,\Lambda^c},\ldots,\omega_{q,\Lambda^c})}
		\nu^{\otimes\Lambda}(d\sigma_\Lambda)=1.
		\]
	Nonnegativity is immediate. Thus
	$A\mapsto\gamma_{q,\Lambda}(A\mid\omega_1,\ldots,\omega_q)$
	is a probability measure on $(\Omega,\mathcal F)$.
	
	Furthermore, the joint measurability assumption on $H_\Lambda$,
	together with the measurability of the maps
	\[
	(\sigma_\Lambda,\omega_i)
	\longmapsto
	\sigma_\Lambda\omega_{i,\Lambda^c},
	\]
	implies that, for every $A\in\mathcal F$,
	\[
	(\omega_1,\ldots,\omega_q)
	\longmapsto
	\gamma_{q,\Lambda}(A\mid\omega_1,\ldots,\omega_q)
	\]
	is $\mathcal F_{\Lambda^c}^{\otimes q}$-measurable. Hence
	$\gamma_{q,\Lambda}$ is a probability $q$-kernel.
	
	\medskip
	
	\noindent
	\textit{Properness.}
	Let $A\in\mathcal F_{\Lambda^c}$. Since $A$ depends only on the
	coordinates outside $\Lambda$, for every
	$\sigma_\Lambda\in\Omega_\Lambda$ and every $i$,
	\[
	\mathbf{1}_A(\sigma_\Lambda\omega_{i,\Lambda^c})
	=
	\mathbf{1}_A(\omega_i).
	\]
	Consequently,
	\[
	\frac{1}{q}\sum_{i=1}^q
	\mathbf{1}_A(\sigma_\Lambda\omega_{i,\Lambda^c})
	=
	\frac{1}{q}\sum_{i=1}^q
	\mathbf{1}_A(\omega_i),
	\]
	and this quantity is independent of $\sigma_\Lambda$. Substitution
	into \eqref{ke} gives
	\[
	\begin{aligned}
		\gamma_{q,\Lambda}(A\mid\omega_1,\ldots,\omega_q)
		&=
		\frac{1}{Z_\Lambda(\omega_1,\ldots,\omega_q)}
		\left(
		\frac{1}{q}\sum_{i=1}^q
		\mathbf{1}_A(\omega_i)
		\right)
		\\
		&\qquad\times
		\int_{\Omega_\Lambda}
		e^{-H_\Lambda(
			\sigma_\Lambda;
			\omega_{1,\Lambda^c},\ldots,\omega_{q,\Lambda^c})}
		\nu^{\otimes\Lambda}(d\sigma_\Lambda)
		=
		\frac{1}{q}\sum_{i=1}^q
		\mathbf{1}_A(\omega_i).
	\end{aligned}
	\]
	Thus $\gamma_{q,\Lambda}$ is proper.
	
	\medskip
	
	\noindent
	\textit{Consistency.}
	The preceding arguments establish that each
	$\gamma_{q,\Lambda}$ is a proper probability $q$-kernel. However,
	the assumptions (1)--(2) alone do not imply the consistency relation
	\eqref{qconsG}. In particular, under the product composition used in
	this paper, the $q$ boundary configurations are resampled
	independently, and the resulting tuple of boundary conditions need
	not have the same empirical boundary distribution as the original
	tuple $(\omega_1,\ldots,\omega_q)$.
	
	Therefore, consistency is an additional compatibility requirement on
	the family of finite-volume Hamiltonians (equivalently, on the
	resulting family of kernels). If \eqref{qconsG} holds, then, by
	definition, the family
	$\gamma=(\gamma_{q,\Lambda})_{\Lambda\Subset V}$ satisfies the
	consistency condition of a $q$-specification. Together with
	properness proved above, this shows that $\gamma$ is a
	$q$-specification.
\end{proof}

\begin{rk}
	For $q=1$, \eqref{ke} becomes
	\[
	\gamma_{1,\Lambda}(A\mid\omega)
	=
	\frac{1}{Z_\Lambda(\omega)}
	\int_{\Omega_\Lambda}
	e^{-H_\Lambda(\sigma_\Lambda;\omega_{\Lambda^c})}
	\mathbf{1}_A(\sigma_\Lambda\omega_{\Lambda^c})
	\,\nu^{\otimes\Lambda}(d\sigma_\Lambda),
	\]
	which is the usual finite-volume Gibbs kernel. When the Hamiltonians
	satisfy the standard compatibility conditions, the corresponding
	family is an ordinary Gibbs specification and the associated
	equilibrium equation is the classical DLR equation.
	
	An important feature of the classical case is that the DLR equation
	is linear in the probability measure $\mu$. Consequently, the set
	$\mathcal G_1(\gamma)$ of Gibbs measures is convex: if
	$\mu_1,\mu_2\in\mathcal G_1(\gamma)$ and $\lambda\in[0,1]$, then
	\[
	\lambda\mu_1+(1-\lambda)\mu_2
	\in
	\mathcal G_1(\gamma).
	\]
	
	This convex structure makes the extremal elements
	$\operatorname{ex}\mathcal G_1(\gamma)$ particularly important; they
	are commonly interpreted as pure phases of the system.
	
	For $q\ge2$, the corresponding equilibrium equation
	\[
	\mu^{\otimes q}\gamma_{q,\Lambda}=\mu
	\]
	is generally nonlinear in $\mu$. Indeed,
	\[
	\bigl(\lambda\mu_1+(1-\lambda)\mu_2\bigr)^{\otimes q}
	\neq
	\lambda\mu_1^{\otimes q}
	+
	(1-\lambda)\mu_2^{\otimes q}
	\]
	in general. Therefore, even when $\mu_1$ and $\mu_2$ are
	$q$-equilibrium measures, their convex combinations need not be
	$q$-equilibrium measures. Thus
	$\mathcal G_q(\gamma)$ need not be convex for $q\ge2$, making its
	structure substantially more delicate than in the classical case.
\end{rk}

\section{Existence of $q$-equilibrium measures}
\label{sec:existence}

Let $\Omega=\Omega_0^V$, where $\Omega_0$ is a finite set and $V$ is a
countable set. We equip $\Omega$ with the product topology. Since
$\Omega_0$ is finite and discrete, a sequence
$(\omega^{(n)})_{n\ge1}$ in $\Omega$ converges to $\omega\in\Omega$ if and
only if, for every $x\in V$,
\[
\omega^{(n)}(x)\to\omega(x), \, n\to\infty.
\]
In particular, $\Omega$ is a compact metrizable space. Consequently,
$\Omega^q$ is also compact and metrizable.

We denote by $\mathcal M_1(\Omega,\mathcal F)$ the space of all
probability measures on $(\Omega,\mathcal F)$, endowed with the weak
topology. Since $\Omega$ is compact metrizable,
$\mathcal M_1(\Omega,\mathcal F)$ is compact in the weak topology.

For a bounded measurable function $f:\Omega\to\mathbb R$ and a
probability $q$-kernel $\gamma_{q,\Lambda}$, we use the notation
\[
\gamma_{q,\Lambda}f(\omega_1,\ldots,\omega_q)
:=
\int_\Omega
f(\sigma)\,
\gamma_{q,\Lambda}
(d\sigma\mid\omega_1,\ldots,\omega_q).
\]

\begin{defn}[Quasilocality]
	A $q$-specification
	$\gamma=(\gamma_{q,\Lambda})_{\Lambda\Subset V}$ is called
	\emph{quasilocal} if, for every finite $\Lambda\Subset V$ and every
	local continuous function $f:\Omega\to\mathbb R$, the map
	\[
	\Omega^q\ni(\omega_1,\ldots,\omega_q)
	\longmapsto
	\gamma_{q,\Lambda}f(\omega_1,\ldots,\omega_q)
	\]
	is continuous.
	
	Since $\Omega_0$ is finite, local continuous functions are finite
	linear combinations of local cylinder indicator functions. Hence,
	equivalently, it is enough to require the above continuity for local
	cylinder indicator functions.
\end{defn}

The next proposition shows that quasilocality, initially formulated for
local functions, extends to all continuous functions on $\Omega$.

\begin{pro}\label{ql-extension}
	Let $\gamma=(\gamma_{q,\Lambda})_{\Lambda\Subset V}$ be a
	quasilocal $q$-specification. Then, for every finite
	$\Lambda\Subset V$ and every $f\in C(\Omega)$, the function
	\[
	\Omega^q\ni(\omega_1,\ldots,\omega_q)
	\longmapsto
	\gamma_{q,\Lambda}f(\omega_1,\ldots,\omega_q)
	\]
	is continuous on $\Omega^q$.
\end{pro}

\begin{proof}
	Fix $\Lambda\Subset V$ and $f\in C(\Omega)$. Since $\Omega$ is a
	compact product of finite discrete spaces, the algebra of local
	continuous functions is uniformly dense in $C(\Omega)$. Hence there
	exists a sequence $(f_m)_{m\ge1}$ of local continuous functions such
	that
	$
	\|f_m-f\|_\infty\longrightarrow0.
	$
	
	For every $m$, quasilocality implies that the function
	\[
	\Omega^q\ni(\omega_1,\ldots,\omega_q)
	\longmapsto
	\gamma_{q,\Lambda}f_m(\omega_1,\ldots,\omega_q)
	\]
	is continuous.
	
	We now show that these functions converge uniformly to
	$\gamma_{q,\Lambda}f$. For every
	$(\omega_1,\ldots,\omega_q)\in\Omega^q$, we have
	\begin{align*}
		&\left|
		\gamma_{q,\Lambda}f(\omega_1,\ldots,\omega_q)
		-
		\gamma_{q,\Lambda}f_m(\omega_1,\ldots,\omega_q)
		\right|
		\\
		&\qquad=
		\left|
		\int_\Omega
		\bigl(f(\sigma)-f_m(\sigma)\bigr)
		\gamma_{q,\Lambda}
		(d\sigma\mid\omega_1,\ldots,\omega_q)
		\right|
		\\
		&\qquad\leq
		\int_\Omega
		|f(\sigma)-f_m(\sigma)|
		\gamma_{q,\Lambda}
		(d\sigma\mid\omega_1,\ldots,\omega_q)
		\leq
		\|f-f_m\|_\infty.
	\end{align*}
	Thus, taking the supremum over
	$(\omega_1,\ldots,\omega_q)\in\Omega^q$, we obtain
	\[
	\sup_{(\omega_1,\ldots,\omega_q)\in\Omega^q}
	\left|
	\gamma_{q,\Lambda}f(\omega_1,\ldots,\omega_q)
	-
	\gamma_{q,\Lambda}f_m(\omega_1,\ldots,\omega_q)
	\right|
	\leq
	\|f-f_m\|_\infty.
	\]
	Therefore, $
	\left\|
	\gamma_{q,\Lambda}f-\gamma_{q,\Lambda}f_m
	\right\|_\infty
	\longrightarrow0.$ 
	Hence $\gamma_{q,\Lambda}f$ is a uniform limit of continuous
	functions on $\Omega^q$, and is consequently continuous.
\end{proof}

For completeness, we next establish the corresponding continuity
property for the product kernel. For a bounded measurable function
$f:\Omega^q\to\mathbb R$, define
\[
\gamma_{q,\Lambda}^{\otimes q}f
(\eta_1,\ldots,\eta_q)
:=
\int_{\Omega^q}
f(\sigma_1,\ldots,\sigma_q)
\prod_{i=1}^q
\gamma_{q,\Lambda}
(d\sigma_i\mid\eta_1,\ldots,\eta_q).
\]
In particular, if
\[
f(\sigma_1,\ldots,\sigma_q)
=
\prod_{i=1}^q f_i(\sigma_i),
\]
where each $f_i:\Omega\to\mathbb R$ is bounded and measurable, then
\[
\gamma_{q,\Lambda}^{\otimes q}f
(\eta_1,\ldots,\eta_q)
=
\prod_{i=1}^q
\gamma_{q,\Lambda}f_i
(\eta_1,\ldots,\eta_q).
\]

\begin{pro}\label{pc}
	Let $\gamma=(\gamma_{q,\Lambda})_{\Lambda\Subset V}$ be a
	quasilocal $q$-specification. Then, for every finite
	$\Lambda\Subset V$ and every $f\in C(\Omega^q)$, the function
	\[
	\Omega^q\ni(\eta_1,\ldots,\eta_q)
	\longmapsto
	\gamma_{q,\Lambda}^{\otimes q}f
	(\eta_1,\ldots,\eta_q)
	\]
	is continuous on $\Omega^q$.
\end{pro}

\begin{proof}
	Fix $\Lambda\Subset V$. We first prove the assertion for a local
	cylinder function $f$ on $\Omega^q$.
	
	Every local cylinder function on $\Omega^q$ depends only on finitely
	many coordinates of each of the $q$ configurations. Since $\Omega_0$
	is finite and discrete, such a function can be written as a finite
	linear combination of product functions of the form
	\[
	f(\sigma_1,\ldots,\sigma_q)
	=
	\prod_{i=1}^q f_i(\sigma_i),
	\]
	where each $f_i:\Omega\to\mathbb R$
	is a local continuous function.
	
	By linearity, it is enough to consider one such product function.
	Using the definition of the product kernel and Fubini's theorem,
	we obtain, for every
	$(\eta_1,\ldots,\eta_q)\in\Omega^q$,
	\begin{align*}
		&\gamma_{q,\Lambda}^{\otimes q}f
		(\eta_1,\ldots,\eta_q)
		=
		\int_{\Omega^q}
		\prod_{i=1}^q f_i(\sigma_i)
		\prod_{i=1}^q
		\gamma_{q,\Lambda}
		(d\sigma_i\mid\eta_1,\ldots,\eta_q)
		\\
		&\quad=
		\prod_{i=1}^q
		\int_\Omega
		f_i(\sigma_i)
		\gamma_{q,\Lambda}
		(d\sigma_i\mid\eta_1,\ldots,\eta_q)
		=
		\prod_{i=1}^q
		\gamma_{q,\Lambda}f_i
		(\eta_1,\ldots,\eta_q).
	\end{align*}
	
	Here, each $f_i$ is a function of the single configuration
	variable $\sigma_i$. After applying the $q$-kernel,
	$\gamma_{q,\Lambda}f_i$ is a function of the $q$ boundary
	configurations $(\eta_1,\ldots,\eta_q)$.
	
	By quasilocality, for every $i=1,\ldots,q$ the map
	\[
	\Omega^q\ni(\eta_1,\ldots,\eta_q)
	\longmapsto
	\gamma_{q,\Lambda}f_i
	(\eta_1,\ldots,\eta_q)
	\]
	is continuous. Therefore, since a finite product of continuous
	functions is continuous,
	\[
	(\eta_1,\ldots,\eta_q)
	\longmapsto
	\gamma_{q,\Lambda}^{\otimes q}f
	(\eta_1,\ldots,\eta_q)
	\]
	is continuous.
	
	Thus $\gamma_{q,\Lambda}^{\otimes q}f$ is continuous for every
	local cylinder function $f$ on $\Omega^q$.
	
	We now extend the result to an arbitrary
	$f\in C(\Omega^q)$. The local cylinder functions form a subalgebra
	of $C(\Omega^q)$, contain the constant functions, and separate
	points of $\Omega^q$. Indeed, suppose that
	$
	(\sigma_1,\ldots,\sigma_q)
	\neq
	(\tau_1,\ldots,\tau_q).
	$
	Then there exist $i\in\{1,\ldots,q\}$ and $x\in V$ such that
	$
	\sigma_i(x)\neq\tau_i(x).
	$
	Because $\Omega_0$ is finite and discrete, there exists a
	one-site cylinder function that distinguishes $\sigma_i$ and
	$\tau_i$. Hence the corresponding cylinder function on $\Omega^q$
	distinguishes
	$(\sigma_1,\ldots,\sigma_q)$ and
	$(\tau_1,\ldots,\tau_q)$.
	
	By the Stone--Weierstrass theorem, local cylinder functions are
	therefore uniformly dense in $C(\Omega^q)$. Thus, for the given
	$f\in C(\Omega^q)$, there exists a sequence of local cylinder
	functions $(f_m)_{m\ge1}$ such that
	$
	\|f_m-f\|_\infty\longrightarrow0.
	$
	
	For every $m$, the function
	\[
	(\eta_1,\ldots,\eta_q)
	\longmapsto
	\gamma_{q,\Lambda}^{\otimes q}f_m
	(\eta_1,\ldots,\eta_q)
	\]
	is continuous.
	
	Moreover, for every
	$(\eta_1,\ldots,\eta_q)\in\Omega^q$,
	\begin{align*}
		&\left|
		\gamma_{q,\Lambda}^{\otimes q}f
		(\eta_1,\ldots,\eta_q)
		-
		\gamma_{q,\Lambda}^{\otimes q}f_m
		(\eta_1,\ldots,\eta_q)
		\right|
		\\
		&\quad=
		\left|
		\int_{\Omega^q}
		\bigl(
		f(\sigma_1,\ldots,\sigma_q)
		-
		f_m(\sigma_1,\ldots,\sigma_q)
		\bigr)
		\prod_{i=1}^q
		\gamma_{q,\Lambda}
		(d\sigma_i\mid\eta_1,\ldots,\eta_q)
		\right|
		\\
		&\quad\leq
		\int_{\Omega^q}
		\left|
		f(\sigma_1,\ldots,\sigma_q)
		-
		f_m(\sigma_1,\ldots,\sigma_q)
		\right|
		\prod_{i=1}^q
		\gamma_{q,\Lambda}
		(d\sigma_i\mid\eta_1,\ldots,\eta_q)
	\leq
		\|f-f_m\|_\infty.
	\end{align*}
	Hence, taking the supremum over
	$(\eta_1,\ldots,\eta_q)\in\Omega^q$, we obtain
	\[
	\sup_{(\eta_1,\ldots,\eta_q)\in\Omega^q}
	\left|
	\gamma_{q,\Lambda}^{\otimes q}f
	(\eta_1,\ldots,\eta_q)
	-
	\gamma_{q,\Lambda}^{\otimes q}f_m
	(\eta_1,\ldots,\eta_q)
	\right|
	\leq
	\|f-f_m\|_\infty.
	\]
	Consequently,
	\[
	\left\|
	\gamma_{q,\Lambda}^{\otimes q}f
	-
	\gamma_{q,\Lambda}^{\otimes q}f_m
	\right\|_\infty
	\longrightarrow0.
	\]
	Thus $\gamma_{q,\Lambda}^{\otimes q}f$ is the uniform limit of
	continuous functions on $\Omega^q$. Therefore,
	$\gamma_{q,\Lambda}^{\otimes q}f$ is continuous.
\end{proof}

We next establish the existence of $q$-equilibrium measures.

\begin{thm}\label{qe}
	If the $q$-specification
	$\gamma=(\gamma_{q,\Lambda})_{\Lambda\Subset V}$ is quasilocal, then
	$	\mathcal G_q(\gamma)\neq\emptyset$.
	\end{thm}
\begin{proof} 	Let
	$
	\Lambda_1\subset\Lambda_2\subset\cdots,
	 \bigcup_{n=1}^\infty\Lambda_n=V,
	$
	be an increasing sequence of finite subsets of $V$. Fix a boundary
	tuple $	(\omega_1,\ldots,\omega_q)\in\Omega^q$
	and define
	\[
	\mu_n(\cdot)
	:=
	\gamma_{q,\Lambda_n}
	(\cdot\mid\omega_1,\ldots,\omega_q).
	\]
	Since $\mathcal M_1(\Omega,\mathcal F)$ is compact in the weak
	topology, there exist a subsequence $(\mu_{n_k})_{k\geq1}$ and a
	probability measure $\mu\in\mathcal M_1(\Omega,\mathcal F)$ such that
	$
	\mu_{n_k}\Longrightarrow\mu.
	$
	
	We show that $\mu\in\mathcal G_q(\gamma)$.
	
	Fix an arbitrary finite set $\Lambda\Subset V$ and
	$f\in C(\Omega)$. Define
	\[
	F_{\Lambda,f}(\eta_1,\ldots,\eta_q)
	:=
	\gamma_{q,\Lambda}f(\eta_1,\ldots,\eta_q)
	=
	\int_\Omega
	f(\sigma)\,
	\gamma_{q,\Lambda}
	(d\sigma\mid\eta_1,\ldots,\eta_q).
	\]
	By Proposition~\ref{ql-extension}, $F_{\Lambda,f}\in C(\Omega^q)$.
	
	Since $\Lambda_n\uparrow V$, we have
	$\Lambda\subseteq\Lambda_{n_k}$ for all sufficiently large $k$.
	By the consistency condition for the $q$-specification,
	\[
	\gamma_{q,\Lambda_{n_k}}
	\gamma_{q,\Lambda}
	=
	\gamma_{q,\Lambda},
	\qquad
	\Lambda\subseteq\Lambda_{n_k}.
	\]
	With the composition rule for $q$-kernels, this gives
	\[
	\begin{aligned}
		\mu_{n_k}^{\otimes q}\gamma_{q,\Lambda}
		&=
		\gamma_{q,\Lambda_{n_k}}^{\otimes q}
		\bigl(\gamma_{q,\Lambda}\bigr)
		(\,\cdot\mid\omega_1,\ldots,\omega_q)
		\\
		&=
		\gamma_{q,\Lambda_{n_k}}
		\gamma_{q,\Lambda}
		(\,\cdot\mid\omega_1,\ldots,\omega_q)
		\\
		&=
		\gamma_{q,\Lambda_{n_k}}
		(\,\cdot\mid\omega_1,\ldots,\omega_q)=
		\mu_{n_k}.
	\end{aligned}
	\]
	Consequently,
	\[
	\begin{aligned}
		\mu_{n_k}(f)=
		\left(
		\mu_{n_k}^{\otimes q}\gamma_{q,\Lambda}
		\right)(f)=
		\mu_{n_k}^{\otimes q}
		\left(
		F_{\Lambda,f}
		\right).
	\end{aligned}
	\]
	
	Because $
	\mu_{n_k}\Longrightarrow\mu
	$
	on the compact metrizable space $\Omega$, finite products preserve weak
	convergence. Hence $\mu_{n_k}^{\otimes q}
	\Longrightarrow
	\mu^{\otimes q}$ on $\Omega^q.$
	Since $F_{\Lambda,f}\in C(\Omega^q)$, we obtain
	\[
	\lim_{k\to\infty}
	\mu_{n_k}^{\otimes q}
	\left(
	F_{\Lambda,f}
	\right)
	=
	\mu^{\otimes q}
	\left(
	F_{\Lambda,f}
	\right).
	\]
	On the other hand, since $f\in C(\Omega)$,
	\[
	\lim_{k\to\infty}\mu_{n_k}(f)
	=
	\mu(f).
	\]
	Therefore,
	\[
	\mu(f)
	=
	\mu^{\otimes q}
	\left(
	F_{\Lambda,f}
	\right).
	\]
	
	Using the definition of $F_{\Lambda,f}$ and Fubini's theorem,
	\[
	\begin{aligned}
		\mu^{\otimes q}
		\left(
		F_{\Lambda,f}
		\right)
		&=
		\int_{\Omega^q}
		\left[
		\int_\Omega
		f(\sigma)\,
		\gamma_{q,\Lambda}
		(d\sigma\mid\eta_1,\ldots,\eta_q)
		\right]
		\mu^{\otimes q}
		(d\eta_1,\ldots,d\eta_q)
		\\
		&=
		\left(
		\mu^{\otimes q}\gamma_{q,\Lambda}
		\right)(f).
	\end{aligned}
	\]
	Hence
	\[
	\mu(f)
	=
	\left(
	\mu^{\otimes q}\gamma_{q,\Lambda}
	\right)(f),
	\qquad
	f\in C(\Omega).
	\]
	Since probability measures on the compact space $\Omega$ are
	determined by their integrals against continuous functions, it follows
	that
		\[
	\mu^{\otimes q}\gamma_{q,\Lambda}
	=
	\mu,
	\qquad
	\Lambda\Subset V.
	\]
	Thus $\mu\in\mathcal G_q(\gamma)$, 	and consequently
	$\mathcal G_q(\gamma)\neq\emptyset$.
\end{proof}

The theorem guarantees existence but does not, in general, imply
uniqueness of the limiting measure. For a fixed boundary tuple
$(\omega_1,\ldots,\omega_q)$, let
$
\mathcal L(\omega_1,\ldots,\omega_q)
$
denote the set of all weak subsequential limits of the finite-volume
measures
\[
\gamma_{q,\Lambda_n}
(\,\cdot\mid\omega_1,\ldots,\omega_q).
\]
The preceding theorem shows that
\[
\mathcal L(\omega_1,\ldots,\omega_q)
\subseteq
\mathcal G_q(\gamma).
\]
In general, $\mathcal L(\omega_1,\ldots,\omega_q)$ may contain more than
one measure. A unique thermodynamic limit can only be asserted under
additional assumptions guaranteeing uniqueness.

If the Gibbs $q$-kernel is symmetric with respect to permutations of the
boundary configurations, then the corresponding finite-volume measures
and their subsequential limits inherit this symmetry. In particular,
for every permutation $\pi$ of $\{1,\ldots,q\}$,
\[
\mathcal L(\omega_{\pi(1)},\ldots,\omega_{\pi(q)})
=
\mathcal L(\omega_1,\ldots,\omega_q).
\]
\begin{pro}
	If $\gamma$ is a quasilocal $q$-specification, then
	$\mathcal G_q(\gamma)$ is a closed subset of
	$\mathcal M_1(\Omega,\mathcal F)$.
\end{pro}

\begin{proof}
	Let $(\mu_n)_{n\geq1}\subset\mathcal G_q(\gamma)$ be a sequence such
	that $\mu_n\Longrightarrow\mu$
	weakly. We show that $\mu\in\mathcal G_q(\gamma)$.
	
	Fix $\Lambda\Subset V$ and $f\in C(\Omega)$. Since
	$\mu_n\in\mathcal G_q(\gamma)$, we have $\mu_n
	=
	\mu_n^{\otimes q}\gamma_{q,\Lambda}.
	$
	Therefore,
	\begin{equation}\label{cc}
		\begin{aligned}
			\int_\Omega f(\sigma)\,\mu_n(d\sigma)
			&=
			\int_\Omega
			f(\sigma)\,
			(\mu_n^{\otimes q}\gamma_{q,\Lambda})(d\sigma)
			\\
			&=
			\int_{\Omega^q}
			F_{\Lambda,f}(\omega_1,\ldots,\omega_q)\,
			\mu_n^{\otimes q}
			(d\omega_1,\ldots,d\omega_q),
		\end{aligned}
	\end{equation}
	where
	\[
	F_{\Lambda,f}(\omega_1,\ldots,\omega_q)
	:=
	\int_\Omega
	f(\sigma)\,
	\gamma_{q,\Lambda}
	(d\sigma\mid\omega_1,\ldots,\omega_q).
	\]
	By Proposition~\ref{ql-extension}, $F_{\Lambda,f}\in C(\Omega^q)$.
		
	Since $	\mu_n\Longrightarrow\mu$
	on the compact metrizable space $\Omega$, finite products preserve weak
	convergence. Hence $\mu_n^{\otimes q}\Longrightarrow
	\mu^{\otimes q}$ on $\Omega^q$.
	
	Consequently, since $F_{\Lambda,f}$ is continuous,
	\[
	\lim_{n\to\infty}
	\int_{\Omega^q}
	F_{\Lambda,f}\,d\mu_n^{\otimes q}
	=
	\int_{\Omega^q}
	F_{\Lambda,f}\,d\mu^{\otimes q}.
	\]
	On the other hand, weak convergence of $\mu_n$ gives
	\[
	\lim_{n\to\infty}
	\int_\Omega f\,d\mu_n
	=
	\int_\Omega f\,d\mu.
	\]
	Combining these identities with \eqref{cc}, we obtain
	\[
	\int_\Omega f\,d\mu
	=
	\int_{\Omega^q}
	F_{\Lambda,f}\,d\mu^{\otimes q}.
	\]
	By the definition of $F_{\Lambda,f}$,
	\[
	\int_{\Omega^q}
	F_{\Lambda,f}\,d\mu^{\otimes q}
	=
	\int_\Omega
	f\,d(\mu^{\otimes q}\gamma_{q,\Lambda}).
	\]
	Thus
	\[
	\int_\Omega f\,d\mu
	=
	\int_\Omega
	f\,d(\mu^{\otimes q}\gamma_{q,\Lambda}),
	\qquad
	f\in C(\Omega).
	\]
	Since probability measures on the compact space $\Omega$ are
	determined by their integrals against continuous functions, it follows
	that
		\[
	\mu^{\otimes q}\gamma_{q,\Lambda}
	=
	\mu,
	\qquad
	\Lambda\Subset V.
	\]
	Therefore $	\mu\in\mathcal G_q(\gamma)$, and hence $\mathcal G_q(\gamma)$ is closed.
\end{proof}

We next describe a standard class of quasilocal Gibbs
$q$-specifications generated by potentials.

\begin{defn}\label{def:potential}
	Let $V$ be equipped with a metric. For each finite set
	$B\Subset V$, let $	\Phi_B:\Omega\to\mathbb R$
	be an $\mathcal F_B$-measurable function, where $\mathcal F_B$
	denotes the $\sigma$-algebra generated by the coordinates in $B$.
	The collection $\Phi=\{\Phi_B\}_{B\Subset V}$
	is called a \emph{potential}.
\end{defn}

Define the \emph{range} of the potential by
\[
r(\Phi)
:=
\inf\left\{
R>0:
\Phi_B\equiv0
\text{ for all }B\Subset V
\text{ with }
\operatorname{diam}(B)>R
\right\}.
\]

For a finite volume $\Lambda\Subset V$, a configuration
$\sigma_\Lambda\in\Omega_\Lambda$, and boundary configurations
$\omega_1,\ldots,\omega_q\in\Omega$, define the finite-volume
Hamiltonian by
\begin{equation}\label{en}
	\begin{aligned}
		H_{\Lambda,\Phi}
		(\sigma_\Lambda;\omega_1,\ldots,\omega_q)
		:={}&
		\sum_{B\Subset\Lambda}
		\Phi_B(\sigma_\Lambda)
		\\
		&+
		\frac{1}{q}
		\sum_{i=1}^q
		\sum_{\substack{
				B\Subset V:\\
				B\cap\Lambda\neq\emptyset,\,
				B\cap\Lambda^c\neq\emptyset
		}}
		\Phi_B
		(\sigma_\Lambda\omega_{i,\Lambda^c}).
	\end{aligned}
\end{equation}
Here, in the first sum, $\Phi_B(\sigma_\Lambda)$ denotes the
evaluation of $\Phi_B$ on the configuration induced by
$\sigma_\Lambda$ on $B\subset\Lambda$.

If $r(\Phi)<\infty$, then $\Phi$ is said to have \emph{finite range}.
In this case, for every finite $\Lambda\Subset V$, only finitely many
terms in \eqref{en} are nonzero.

If $r(\Phi)=\infty$, then $\Phi$ has \emph{infinite range}. In this
case, we assume that $\Phi$ is absolutely summable, namely,
\begin{equation}\label{eq:abs-summable}
	\sum_{\substack{B\Subset V\\B\ni x}}
	\|\Phi_B\|_\infty
	<\infty,
	\qquad
	x\in V,
\end{equation}
where
\[
\|\Phi_B\|_\infty
:=
\sup_{\omega\in\Omega}|\Phi_B(\omega)|.
\]
In particular, \eqref{eq:abs-summable} guarantees that
$\|\Phi_B\|_\infty<\infty$ for every $B$ and that the Hamiltonian
\eqref{en} is well defined.

\begin{pro}\label{sc}
	Let $\Phi$ be a potential of finite range or an absolutely summable
	potential. Then the Gibbs $q$-kernel $\gamma^\Phi$, defined by
	\eqref{ke} with the Hamiltonian \eqref{en}, is quasilocal.
\end{pro}

\begin{proof}
	Fix a finite volume $\Lambda\Subset V$ and let
	\[
	\omega=(\omega_1,\ldots,\omega_q),
	\qquad
	\omega'=(\omega'_1,\ldots,\omega'_q)
	\]
	be two tuples of boundary configurations. Let
	$\Delta\Subset V$ be a finite set with $\Lambda\subseteq\Delta$ and
	suppose that
	\[
	\omega_i|_\Delta=\omega'_i|_\Delta,
	\qquad
	i=1,\ldots,q.
	\]
	Put
	\[
	D:=\operatorname{dist}(\Lambda,\Delta^c).
	\]
	
	For every $\eta_\Lambda\in\Omega_\Lambda$, all terms in
	\eqref{en} corresponding to sets $B\subseteq\Delta$ are identical
	for $\omega$ and $\omega'$. Hence only sets satisfying
	\[
	B\cap\Lambda\neq\emptyset,
	\qquad
	B\cap\Delta^c\neq\emptyset
	\]
	can contribute to the difference. For such sets we have 
	\[
	\operatorname{diam}(B)
	\geq
	\operatorname{dist}(\Lambda,\Delta^c)
	=
	D.
	\]

	Consequently,
	\[
	\begin{aligned}
		&\left|
		H_{\Lambda,\Phi}
		(\eta_\Lambda;\omega)
		-
		H_{\Lambda,\Phi}
		(\eta_\Lambda;\omega')
		\right|
		\\
		&\qquad\leq
		\frac{1}{q}
		\sum_{i=1}^q
		\sum_{\substack{
				B\cap\Lambda\neq\emptyset\\
				B\cap\Delta^c\neq\emptyset
		}}
		\left|
		\Phi_B(\eta_\Lambda\omega_{i,\Lambda^c})
		-
		\Phi_B(\eta_\Lambda\omega'_{i,\Lambda^c})
		\right|
		\\
		&\qquad\leq
		2
		\sum_{\substack{
				B\cap\Lambda\neq\emptyset\\
				B\cap\Delta^c\neq\emptyset
		}}
		\|\Phi_B\|_\infty.
	\end{aligned}
	\]

	Moreover, every such $B$ contains at least one point $x\in\Lambda$.
	Therefore,
	\[
	\begin{aligned}
		&\sup_{\eta_\Lambda\in\Omega_\Lambda}
		\left|
		H_{\Lambda,\Phi}
		(\eta_\Lambda;\omega)
		-
		H_{\Lambda,\Phi}
		(\eta_\Lambda;\omega')
		\right|
		\\
		&\qquad\leq
		2|\Lambda|
		\max_{x\in\Lambda}
		\sum_{\substack{
				B\ni x\\
				\operatorname{diam}(B)\geq D
		}}
		\|\Phi_B\|_\infty.
	\end{aligned}
	\]
	For an absolutely summable potential, the right-hand side tends to zero
	as $D\to\infty$. For a finite-range potential, the right-hand side is
	zero once $D>r(\Phi)$.
	
	Set
	\[
	\delta_D:=
	\sup_{\eta_\Lambda\in\Omega_\Lambda}
	\left|
	H_{\Lambda,\Phi}(\eta_\Lambda;\omega)
	-
	H_{\Lambda,\Phi}(\eta_\Lambda;\omega')
	\right|.
	\]
	Then $\delta_D\to0$ as $D\to\infty$.
	
	Since $\Omega_\Lambda$ is finite, the Gibbs probabilities depend
	continuously on the corresponding finite-volume Hamiltonian.
	Therefore,
	\[
	\sup_{\tau_\Lambda\in\Omega_\Lambda}
	\left|
	\gamma^\Phi_{q,\Lambda}
	(\tau_\Lambda\mid\omega)
	-
	\gamma^\Phi_{q,\Lambda}
	(\tau_\Lambda\mid\omega')
	\right|
	\longrightarrow0
	\qquad\text{as }D\to\infty.
	\]
	Consequently, for every local continuous function
	$f:\Omega\to\mathbb R$,
	\[
	\left|
	\gamma^\Phi_{q,\Lambda}f(\omega)
	-
	\gamma^\Phi_{q,\Lambda}f(\omega')
	\right|
	\longrightarrow0
	\qquad\text{as }D\to\infty.
	\]
	Thus
	\[
	(\omega_1,\ldots,\omega_q)
	\longmapsto
	\gamma^\Phi_{q,\Lambda}f(\omega_1,\ldots,\omega_q)
	\]
	is continuous on $\Omega^q$. Hence $\gamma^\Phi$ is quasilocal.
\end{proof}

\begin{rk}
	Proposition~\ref{sc} gives a standard method for constructing
	quasilocal Gibbs $q$-kernels from finite-range or absolutely summable
	potentials. Combined with Theorem~\ref{qe}, whenever these kernels
	form a $q$-specification, it yields the existence of
	$q$-equilibrium measures.
\end{rk}

\section{Uniqueness via sensitivity to boundary conditions}

 A function $ f : \Omega^q \to \mathbb{R} $ is local if there exists a finite subset $ \Lambda \subset V $ such that for any $ (\omega_1, \dots, \omega_q), (\omega_1', \dots, \omega_q') \in \Omega^q $, if $ \omega_i|_\Lambda = \omega_i'|_\Lambda $ for all $ i = 1, \dots, q $, then $ f(\omega_1, \dots, \omega_q) = f(\omega_1', \dots, \omega_q') $.
In other words, a local function depends only on the configurations within a finite set $ \Lambda $. Such functions are bounded and measurable.

The aim of this section is to prove the following result, gives a criterion for uniqueness of a $q$-equilibrium measure.

\begin{pro}
The following are equivalent
\begin{enumerate}
    \item Uniqueness holds: $\mathcal{G}_q(\gamma) = \{\mu\}$;
    \item For all $\omega_1, \dots, \omega_q$, all $\Lambda_n \uparrow V$ and all local functions $f$,
    $$
    (\gamma^{\otimes q}_{\Lambda_n}f)(\omega_1, \dots, \omega_q) \to \mu^{\otimes q}(f). 
    $$
\end{enumerate}
\end{pro}
\begin{proof}	($1 \Rightarrow 2$): Let $\omega_1, \dots, \omega_q$ be a boundary condition. By the proof of Theorem \ref{qe} we know that, from any sequence $(\gamma_{\Lambda_n}(\cdot|\omega_1,\dots, \omega_q))_{n\geq 1}$, one can extract a subsequence converging to some element of $\mathcal{G}_q(\gamma)$. If $\mathcal{G}_q(\gamma) = \{\mu\}$, all these subsequences must have the same limit $\mu$. Therefore, the sequence itself converges to $\mu$. Then for any local function $f$ on $\Omega^q$, we have
	$$
	(\gamma^{\otimes q}_{\Lambda_n} f)(\omega_1, \dots, \omega_q) = \int_{\Omega^q} f(\sigma_1, \dots, \sigma_q) \prod_{i=1}^q \gamma_{\Lambda_n}(d\sigma_i|\omega_1,\dots,\omega_q)$$ $$ \to \int_{\Omega^q} f(\sigma_1, \dots, \sigma_q) \prod_{i=1}^q \mu(d\sigma_i) = \mu^{\otimes q}(f).
	$$
	
	($2 \Rightarrow 1$): Assume condition 2 holds. Let $\nu \in \mathcal{G}_q(\gamma)$ be any $q$-equilibrium measure. We will show that $\nu = \mu$. 
	
	Let $g$ be any local function on $\Omega$. Define a function $f$ on $\Omega^q$ by $f(\sigma_1, \dots, \sigma_q) = g(\sigma_1)$. Then $f$ is local. By condition 2, for every $\omega_1, \dots, \omega_q$, we have
	$$
	(\gamma^{\otimes q}_{\Lambda_n} f)(\omega_1, \dots, \omega_q) \to \mu^{\otimes q}(f).
	$$
	Taking into account
	$$
	(\gamma^{\otimes q}_{\Lambda_n}) f(\omega_1, \dots, \omega_q) = \int_{\Omega^q} g(\sigma_1) \prod_{i=1}^q \gamma_{\Lambda_n}(d\sigma_i|\omega_1,\dots,\omega_q) = \gamma_{\Lambda_n}(g \mid \omega_1, \dots, \omega_q),
	$$
	and
	$$
	\mu^{\otimes q}(f) = \int_{\Omega^q} g(\sigma_1) \prod_{i=1}^q \mu(d\sigma_i) = \mu(g),
	$$
	we get
	$$
	\gamma_{\Lambda_n}(g \mid \omega_1, \dots, \omega_q) \to \mu(g) \quad \text{for all } \omega_1, \dots, \omega_q.
	$$
	
	Now, since $\nu$ is $q$-equilibrium,  the equation \eqref{DLR} implies
	$$
	\nu(g) = (\nu^{\otimes q} \gamma_{\Lambda_n})(g) = \int_{\Omega^q} \gamma_{\Lambda_n}(g \mid \omega_1, \dots, \omega_q) \prod_{i=1}^q\nu(d\omega_i).
	$$
	Since $\gamma_{\Lambda_n}(g \mid \omega_1, \dots, \omega_q)$ is bounded (as $g$ is bounded) and converges pointwise to $\mu(g)$, by the dominated convergence theorem:
	$$
	\nu(g) = \lim_{n\to\infty} \int_{\Omega^q} \gamma_{\Lambda_n}(g \mid \omega_1, \dots, \omega_q) \prod_{i=1}^q\nu(d\omega_i) = \mu(g).
	$$
	Since this holds for every local function $g$, we conclude that $\nu = \mu$. Therefore, $\mathcal{G}_q(\gamma) = \{\mu\}$.
\end{proof}

\section{A $q$-specification with empty-set of $q$-equilibrium measures} Here we generalize the result of \cite[Exersize 6.15]{FV} to the case $q\geq 1$.
Consider the configuration space $\Omega = \{-1, +1\}^V$, 
where $V$ is a countable set. Define the following configurations:

$\sigma^-$: all spins are $-1$.

$\sigma^{-,x}$: all spins are $-1$ except at site $x$, where the spin is $+1$.

For $\Lambda \Subset V$ and $\omega_1, \dots, \omega_q \in \Omega$, define
\begin{equation}\label{qs0}   
\gamma_\Lambda(A \mid \omega_1, \dots, \omega_q) :={1\over q}
\begin{cases}
\displaystyle \frac{1}{|\Lambda|} \sum_{x \in \Lambda} \sum_{i=1}^q \mathbf{1}_A(\sigma_\Lambda^{-,x} \omega_{i,\Lambda^c}), & \text{if } \omega_{i,\Lambda^c} = \sigma^-_{\Lambda^c} \text{ for some } i, \\
\displaystyle \sum_{i=1}^q \mathbf{1}_A(\sigma^-_\Lambda \omega_{i,\Lambda^c}), & \text{otherwise}.
\end{cases}
 \end{equation}

\begin{pro} The $\gamma=\{\gamma_\Lambda\}$, defined in (\ref{qs0}), is a $q$-specification and  $\mathcal G_q(\gamma) = \emptyset$.
\end{pro}
\begin{proof}  Let us verify that $\gamma$ is a $q$-specification.

{\it Properness}:
Let $A \in \mathcal{F}_{\Lambda^c}$. Then for any configuration $\xi$, the value of $\mathbf{1}_A(\xi)$ depends only on $\xi_{\Lambda^c}$. Since $\sigma_\Lambda^{-,x}$ and $\sigma^-_\Lambda$ affect only $\Lambda$, we have
$$
\mathbf{1}_A(\sigma_\Lambda^{-,x} \omega_{i,\Lambda^c}) = \mathbf{1}_A(\omega_i), \quad
\mathbf{1}_A(\sigma^-_\Lambda \omega_{i,\Lambda^c}) = \mathbf{1}_A(\omega_i).
$$
Thus, in both cases we get
$$
\gamma_\Lambda(A \mid \omega_1, \dots, \omega_q) = {1\over q}\sum_{j=1}^q \mathbf{1}_A(\omega_j).
$$
Hence $\gamma_\Lambda$ is proper.

{\it Consistency}: Let $\Lambda \subset \Delta \Subset V$. To verify that
$$
(\gamma_\Delta \gamma_\Lambda)(A \mid \eta_1, \dots, \eta_q) = \gamma_\Delta(A \mid \eta_1, \dots, \eta_q)
$$
for all $A \in \mathcal{F}$ and $\eta_1, \dots, \eta_q \in \Omega$, we consider two cases based on the boundary conditions $\eta_1, \dots, \eta_q$ on $\Delta^c$.

Case 1: There exists some $j$ such that $\eta_{j,\Delta^c} = \sigma^-_{\Delta^c}$.

For any $\zeta_i$ in the support of $\gamma_\Lambda(\cdot \mid \eta_1, \dots, \eta_q)$, by properness, $\zeta_{i,\Lambda^c} = \eta_{i,\Lambda^c}$. Since $\Delta^c \subset \Lambda^c$, we have $\zeta_{i,\Delta^c} = \eta_{i,\Delta^c}$. In particular, for the index $j$ with $\eta_{j,\Delta^c} = \sigma^-_{\Delta^c}$, we have $\zeta_{j,\Delta^c} = \sigma^-_{\Delta^c}$. Therefore
$$
\gamma_\Delta(A \mid \zeta_1, \dots, \zeta_q) = \frac{1}{q|\Delta|} \sum_{k \in \Delta} \sum_{j=1}^q \mathbf{1}_A(\sigma_\Delta^{-,k} \zeta_{j,\Delta^c}) = \frac{1}{q|\Delta|} \sum_{k \in \Delta} \sum_{j=1}^q \mathbf{1}_A(\sigma_\Delta^{-,k} \eta_{j,\Delta^c}).
$$

Thus,
$$
(\gamma_\Delta \gamma_\Lambda)(A \mid \eta_1, \dots, \eta_q) = \int \left( \frac{1}{q|\Delta|} \sum_{k \in \Delta} \sum_{j=1}^q \mathbf{1}_A(\sigma_\Delta^{-,k} \eta_{j,\Delta^c}) \right) \prod_{i=1}^q \gamma_\Lambda(d\zeta_i \mid \eta_1, \dots, \eta_q) 
$$ $$= \frac{1}{q|\Delta|} \sum_{k \in \Delta} \sum_{j=1}^q \mathbf{1}_A(\sigma_\Delta^{-,k} \eta_{j,\Delta^c})=\gamma_\Delta(A \mid \eta_1, \dots, \eta_q),
$$
since the integrand is constant and the measures integrate to 1. 

Case 2: For all $j$, $\eta_{j,\Delta^c} \neq \sigma^-_{\Delta^c}$.


Again, for any $\zeta_i$ in the support, $\zeta_{i,\Delta^c} = \eta_{i,\Delta^c}$, so $\zeta_{i,\Delta^c} \neq \sigma^-_{\Delta^c}$ for all $i$. Hence,
$$
\gamma_\Delta(A \mid \zeta_1, \dots, \zeta_q) = {1\over q}\sum_{j=1}^q \mathbf{1}_A(\sigma^-_\Delta \zeta_{j,\Delta^c}) = {1\over q}\sum_{j=1}^q \mathbf{1}_A(\sigma^-_\Delta \eta_{j,\Delta^c}).
$$

Integrating gives
$$
(\gamma_\Delta \gamma_\Lambda)(A \mid \eta_1, \dots, \eta_q) = {1\over q}\sum_{j=1}^q \mathbf{1}_A(\sigma^-_\Delta \eta_{j,\Delta^c})=\gamma_\Delta(A \mid \eta_1, \dots, \eta_q).
$$

Therefore, $\gamma_\Delta \gamma_\Lambda = \gamma_\Delta$, and $\gamma$ is consistent.

{\it Nonexistence of $q$-equilibrium measures}: 
Recall that a probability measure $\mu$ on $\Omega = \{-1, +1\}^V$ is a $ q $-equilibrium measure
for the $ q $-specification $\gamma$ if for every finite $\Lambda \Subset V$ and every $A \in \mathcal{F}$, 
the following holds:
\begin{equation}\label{s0}
\mu(A) = \int_{\Omega^q} \gamma_\Lambda(A \mid \omega_1, \dots, \omega_q) \, \prod_{i=1}^q\mu(d\omega_i).
\end{equation}

We prove that no such non-zero measure $\mu$ exists, for $q$-specification (\ref{qs0}).

For any finite $\Lambda \Subset V$, define the event:
$$
A_\Lambda = \{ \omega \in \Omega: \omega_\Lambda = \sigma^-_\Lambda \}.
$$

Consider two cases:

Case 1: There exists some $i$ such that $\omega_{i,\Lambda^c} = \sigma^-_{\Lambda^c}$.  
  Then  
  $$
  \gamma_\Lambda(A_\Lambda \mid \omega_1, \dots, \omega_q) = \frac{1}{q|\Lambda|} \sum_{x \in \Lambda} \sum_{i=1}^q \mathbf{1}_{A_\Lambda}(\sigma_\Lambda^{-,x} \omega_{i,\Lambda^c}).
  $$  
  For any $x \in \Lambda$, the configuration $\sigma_\Lambda^{-,x} \omega_{i,\Lambda^c}$ has a $+1$ at $x$ and $-1$ on $\Lambda \setminus \{x\}$, so it is not in $A_\Lambda$. Hence, each indicator is 0, and the sum is 0.

Case 2: For all $i$, $\omega_{i,\Lambda^c} \neq \sigma^-_{\Lambda^c}$.  
  Then  
  $$
  \gamma_\Lambda(A_\Lambda \mid \omega_1, \dots, \omega_q) ={1\over q} \sum_{i=1}^q \mathbf{1}_{A_\Lambda}(\sigma^-_\Lambda \omega_{i,\Lambda^c}).
  $$  
  The configuration $\sigma^-_\Lambda \omega_{i,\Lambda^c}$ is $-1$ on $\Lambda$ and equals $\omega_{i,\Lambda^c}$ on $\Lambda^c$, so it is in $A_\Lambda$. Hence, each indicator is 1, and the sum is 1.

Therefore,
$$
\gamma_\Lambda(A_\Lambda \mid \omega_1, \dots, \omega_q) = 
\begin{cases}
0, & \text{if } \, \omega_{i,\Lambda^c} = \sigma^-_{\Lambda^c}, \ \text{for some} \, i \\
1, & \text{otherwise}.
\end{cases}
$$

Thus, the right-hand side of the equation \eqref{s0} becomes
$$\int_{\Omega^q} \gamma_\Lambda(A_\Lambda \mid \omega_1, \dots, \omega_q) \, \prod_{i=1}^q\mu(d\omega_i)
  = \left( \mu( \{ \omega : \omega_{\Lambda^c} \neq \sigma^-_{\Lambda^c} \} ) \right)^q.
$$

Let $p_\Lambda = \mu( \{ \omega : \omega_{\Lambda^c} = \sigma^-_{\Lambda^c} \} )$. Then $\mu( \{ \omega : \omega_{\Lambda^c} \neq \sigma^-_{\Lambda^c} \} ) = 1 - p_\Lambda$, and the equation \eqref{s0} gives:
\begin{equation}\label{hd}
\mu(A_\Lambda) = (1 - p_\Lambda)^q. 
\end{equation}
Now consider an increasing sequence of finite sets $\Lambda_n \uparrow V$. Then
$$\bigcap_{n\geq 1}\{ \omega : \omega_{\Lambda_n^c} = \sigma^-_{\Lambda_n^c} \} = \{ \sigma^- \},$$  
  so by continuity (from above) of measure,  
  $$
  \lim_{n \to \infty} p_{\Lambda_n} = \mu(\{\sigma^-\}).
  $$
 In the other hand $\{ A_{\Lambda_n} \} = \{ \omega : \omega_{\Lambda_n} = \sigma^-_{\Lambda_n} \} \uparrow \{ \sigma^- \}$,  
  so  by continuity (from below) of measure
  $$
  \lim_{n \to \infty} \mu(A_{\Lambda_n}) = \mu(\{\sigma^-\}).
  $$
Taking the limit in equation (\ref{hd}) as $n \to \infty$, we obtain
\begin{equation}\label{oo}
\mu(\{\sigma^-\}) = (1 - \mu(\{\sigma^-\}))^q. 
\end{equation}
  
Now apply the equation (\ref{s0}) to the singleton $A = \{\sigma^-\}$. Compute $\gamma_\Lambda(\{\sigma^-\} \mid \omega_1, \dots, \omega_q)$ based on the boundary conditions:

- If there exists some $i$ such that $\omega_{i,\Lambda^c} = \sigma^-_{\Lambda^c}$, then
  $$
  \gamma_\Lambda(\{\sigma^-\} \mid \omega_1, \dots, \omega_q) = \frac{1}{q|\Lambda|} \sum_{x \in \Lambda} \sum_{i=1}^q \mathbf{1}_{\{\sigma^-\}}(\sigma_\Lambda^{-,x} \omega_{i,\Lambda^c}).
  $$
  For any $x \in \Lambda$, the configuration $\sigma_\Lambda^{-,x} \omega_{i,\Lambda^c}$ has a $+1$ at $x$ and $-1$ elsewhere, so it is not equal to $\sigma^-$. Hence, each indicator is 0, and the sum is 0.

- If for all $i$, $\omega_{i,\Lambda^c} \neq \sigma^-_{\Lambda^c}$, then
  $$
  \gamma_\Lambda(\{\sigma^-\} \mid \omega_1, \dots, \omega_q) = {1\over q}\sum_{i=1}^q \mathbf{1}_{\{\sigma^-\}}(\sigma^-_\Lambda \omega_{i,\Lambda^c}).
  $$
  The configuration $\sigma^-_\Lambda \omega_{i,\Lambda^c}$ is $-1$ on $\Lambda$ and equals $\omega_{i,\Lambda^c}$ on $\Lambda^c$. For this to equal $\sigma^-$, we must have $\omega_{i,\Lambda^c} = \sigma^-_{\Lambda^c}$ for some $i$, which contradicts the assumption. Hence, all indicators are 0, and the sum is 0.

Thus, in all cases, $\gamma_\Lambda(\{\sigma^-\} \mid \omega_1, \dots, \omega_q) = 0$, so by (\ref{s0}) we get 
$\mu(\{\sigma^-\}) = 0$. Substituting into equation (\ref{oo}) yields: $0 = (1 - 0)^q = 1$, a contradiction.

Hence, the assumption that $\mu \in \mathcal{G}_q(\gamma)$ leads to a contradiction. 
Therefore, no such measure exists, and $\mathcal{G}_q(\gamma) = \emptyset$.
\end{proof} 

\section{Dynamical systems generated by $q$-SO and $q$-equilibrium measures}

As shown above, each $q$-equilibrium measure is a fixed point of a $q$-SO. In this section, we study the dynamical systems 
generated by such $q$-SOs. Any limit point of the trajectories of these systems 
is therefore a $q$-equilibrium measure. 

\medskip

For each $\Lambda \Subset V$, the $q$-SO $V_{q,\Lambda}$ acting on the set $ \mathcal M_1(\Omega, \mathcal F)$ 
is defined as follows (see (\ref{Vq})): for an arbitrary measure $\lambda \in \mathcal M_1(\Omega, \mathcal F)$, the measure $\lambda' = V_{q, \Lambda}(\lambda)$ is given by
\begin{equation}\label{Vqu}
	V_{q,\Lambda}(\lambda)(A) := \lambda'(A)
	= \int_{\Omega^q} \gamma_\Lambda(A \mid \sigma_{1},\dots, \sigma_{q}) \prod_{i=1}^q \lambda(d\sigma_i),
	\quad A \in \mathcal F.
\end{equation}

For a given initial measure $\lambda^{(0)} = \lambda$, we define the trajectory of the operator~(\ref{Vqu}) by
$$
\lambda^{(n+1)}(A) = V_{q,\Lambda}(\lambda^{(n)})(A), \qquad n = 0,1,2,\dots
$$
The sequence $\{\lambda^{(n)}\}_{n \ge 0}$ represents the discrete-time dynamical system generated by $V_{q,\Lambda}$. 
Its asymptotic behavior--particularly its fixed points and limit measures--reveals the structure of $q$-equilibrium states.

\medskip
In general, the explicit computation of iterates of $V_{q,\Lambda}$ for a given $q$-specification $\gamma_\Lambda$ is difficult. 
To make progress, we consider a class of $q$-specifications for which such computations are tractable.

{\bf Assumption 1.} {\it The $q$-kernel $\gamma_\Lambda$ depends only on the empirical measure 
	of its conditioning variables, i.e., there exists a measurable mapping
	$\tilde{\gamma}_\Lambda : \mathcal F \times \mathcal M_1(\Omega) \to [0,1]$ such that
	$$
	\gamma_\Lambda(A \mid \sigma_1, \dots, \sigma_q)
	= \tilde{\gamma}_\Lambda\!\left(A; \frac{1}{q} \sum_{i=1}^q \delta_{\sigma_i}\right),
	$$
	where $\delta_{\sigma}$ denotes the Dirac measure at $\sigma$.} 
	
	From this assumption it follows also that  $\gamma_\Lambda$ is \emph{symmetric} in its conditioning arguments: for any permutation $\pi$ of $\{1,\dots,q\}$ and any $A \in \mathcal{F}$,
	$$
	\gamma_\Lambda(A \mid \sigma_1, \dots, \sigma_q)
	= \gamma_\Lambda(A \mid \sigma_{\pi(1)}, \dots, \sigma_{\pi(q)}).
	$$

\begin{lemma}\label{VV}
	If Assumption 1 is satisfied, then for any fixed $\Lambda \Subset V$, we have
	$$
	V_{q,\Lambda}^2 = V_{q,\Lambda}.
	$$
\end{lemma}

\begin{proof}
	We compute the second iterate of $V_{q,\Lambda}$. By definition, we have
	$$
	V_{q,\Lambda}^2(\lambda)(A)
	= \int_{\Omega^q} \gamma_\Lambda(A \mid \zeta_1, \dots, \zeta_q)
	\prod_{i=1}^q V_{q,\Lambda}(\lambda)(d\zeta_i).
	$$
	Each $V_{q,\Lambda}(\lambda)(d\zeta_i)$ can be expanded using~(\ref{Vqu}):
	$$
	V_{q,\Lambda}(\lambda)(d\zeta_i)
	= \int_{\Omega^q} \gamma_\Lambda(d\zeta_i \mid \sigma^{(i)}_1, \dots, \sigma^{(i)}_q)
	\prod_{j=1}^q \lambda(d\sigma^{(i)}_j).
	$$
	Substituting this into the first expression and applying Fubini's theorem, we obtain
	\begin{equation}\label{eq:V2}
		\begin{aligned}
			V_{q,\Lambda}^2(\lambda)(A)
			&= \int_{\Omega^{q^2}} \left[
			\int_{\Omega^q} \gamma_\Lambda(A \mid \zeta_1, \dots, \zeta_q)
			\prod_{i=1}^q \gamma_\Lambda(d\zeta_i \mid \sigma^{(i)}_1, \dots, \sigma^{(i)}_q)
			\right] \prod_{i=1}^q \prod_{j=1}^q \lambda(d\sigma^{(i)}_j).
		\end{aligned}
	\end{equation}
	
	Let us denote the inner integral by
	$$
	I(A; \{\sigma^{(i)}_j\})
	:= \int_{\Omega^q} \gamma_\Lambda(A \mid \zeta_1, \dots, \zeta_q)
	\prod_{i=1}^q \gamma_\Lambda(d\zeta_i \mid \sigma^{(i)}_1, \dots, \sigma^{(i)}_q).
	$$
	Under Assumption 1, $I(A; \{\sigma^{(i)}_j\})$
	depends only on the \emph{empirical measure}
	$$
	L = \frac{1}{q^2} \sum_{i=1}^q \sum_{j=1}^q \delta_{\sigma^{(i)}_j}.
	$$
	Indeed, since each $\gamma_\Lambda(d\zeta_i \mid \sigma^{(i)}_1, \dots, \sigma^{(i)}_q)$
	depends only on the empirical distribution
	$\frac{1}{q} \sum_{j=1}^q \delta_{\sigma^{(i)}_j}$,
	and $\gamma_\Lambda(A \mid \zeta_1, \dots, \zeta_q)$ depends only on
	$\frac{1}{q} \sum_{i=1}^q \delta_{\zeta_i}$,
	the double composition preserves dependence on the combined empirical measure $L$.
	Moreover, by the exchangeability of the product measure
	$\prod_{i,j} \lambda(d\sigma^{(i)}_j)$,
	the order of variables $\{\sigma^{(i)}_j\}$ is irrelevant.
	
	Hence, one can symmetrize the variables and replace the $q^2$ spins
	$\{\sigma^{(i)}_j\}$ by $q$ i.i.d. variables
	$\sigma_1, \dots, \sigma_q$ drawn from $\lambda$,
	without changing the value of the integral.
		
	Consequently, the inner integral can be rewritten as (see Lemma \ref{ggg})
	$$
	I(A; \{\sigma^{(i)}_j\})=\gamma_\Lambda\gamma_\Lambda(A \mid \sigma_1, \dots, \sigma_q)
	= \gamma_\Lambda(A \mid \sigma_1, \dots, \sigma_q),
	$$
	and the product measure $\prod_{i,j} \lambda(d\sigma^{(i)}_j)$
	reduces to $\prod_{i=1}^q \lambda(d\sigma_i)$.
	Substituting this back into~(\ref{eq:V2}), we obtain
	$$
	V_{q,\Lambda}^2(\lambda)(A)
	= \int_{\Omega^q} \gamma_\Lambda(A \mid \sigma_1, \dots, \sigma_q)
	\prod_{i=1}^q \lambda(d\sigma_i)
	= V_{q,\Lambda}(\lambda)(A).
	$$
	Hence $V_{q,\Lambda}^2 = V_{q,\Lambda}$, as claimed.
\end{proof}

\medskip
By Lemma~\ref{VV}, under Assumption 1, each image $V_{q,\Lambda}(\mu)$ is a fixed point of $V_{q,\Lambda}$, i.e.,
$$
V_{q,\Lambda}(V_{q,\Lambda}(\mu)) = V_{q,\Lambda}(\mu),
\quad \text{for all } \mu \in \mathcal M_1(\Omega, \mathcal F).
$$
Thus, every measure in the image of $V_{q,\Lambda}$ is a $q$-equilibrium measure
corresponding to $\gamma_\Lambda$, i.e., we have proved the following

\begin{thm}\label{thm:Gq-image}
	If Assumption 1 is satisfied, then
	$$
	{\mathcal G}_q(\gamma)
	= \bigcap_{\Lambda\Subset V} {\rm Im}(V_{q,\Lambda}).
	$$
	\end{thm}

\section{1D Ising model with multiple $q$-equilibrium measures}

Let $V = \mathbb{Z}$, spin space $\Omega_0 = \{\pm1\}$, and $\Omega = \Omega_0^V$. Fix integers $q \ge 1$ and a finite volume $\Lambda \Subset V$. Let $J, h\in \mathbb{R}$  and define 
\begin{equation}\label{1dh}  H(\sigma_\Lambda; \omega_1, \dots, \omega_q) = -J \sum_{\substack{\{x,y\} \subset \Lambda \\ |x-y|=1}} \sigma_x \sigma_y - h \sum_{x \in \Lambda} \sigma_x - J \sum_{\substack{x \in \Lambda,\, y \in \Lambda^c \\ |x-y|=1}} \sigma_x \, \overline{\omega}_y, \end{equation}
where $\sigma_x\in \Omega_0$, $$ \overline{\omega}_y = \frac{1}{q} \sum_{i=1}^q \omega_{i,y}. $$ 

 We are going to prove the following 
\begin{thm}\label{tt3} For 1D model (\ref{1dh}) if $q\geq 2$, $J>0$, $h=0$ then there is $J_*>0$ such that for each temperature $T \in \left(0, {J\over J_*}\right)$ there are at least three translation-invariant $q$-equilibrium measures.
\end{thm}
\begin{proof} 
{\it The single-site equation}:   Fix the origin $0 \in V$ and take $\Lambda = \{0\}$.
The restriction of the Hamiltonian is
$$
H(\sigma_0; \omega_1, \dots, \omega_q) = -h \sigma_0 - J \sigma_0 \sum_{y \in \partial\{0\}} \overline{\omega}_y,
$$
where $ \partial\{0\} = \{-1, +1\} $. 

Define the boundary function as
$$
\mathbf{b}(\omega_1, \dots, \omega_q) = \sum_{y \in \partial\{0\}} \overline{\omega}_y = \frac{1}{q} \sum_{i=1}^q (\omega_{i,-1} + \omega_{i,1}).
$$

The single-site $ q $-kernel (at inverse temperature $ \beta $) gives 
\begin{equation}\label{ga}
\gamma_{\{0\}}(\sigma_0 = +1 \mid \omega_1, \dots, \omega_q) = \frac{e^{\beta(h + J \mathbf{b})}}{e^{\beta(h + J \mathbf{b})} + e^{-\beta(h + J \mathbf{b})}} = \frac{1 + \tanh(\beta(h + J \mathbf{b}))}{2}.
\end{equation}

{\it Self-consistency equation}:
By compatibility of $\mu$ with the $ q $-specification $ \gamma $, for $ \Lambda = \{0\} $ and the event $ A = \{\sigma : \sigma_0 = +1\} $, we have

$$
\mu(\sigma_0 = +1) = \int_{\Omega^q} \gamma_{\{0\}}(\sigma_0 = +1 \mid \omega_1, \dots, \omega_q) \prod_{i=1}^q \mu(d\omega_i).
$$

Substituting $\gamma_{\{0\}}$ given by (\ref{ga}) we obtain
$$
\mu(\sigma_0 = +1) = \int_{\Omega^q} \frac{1 + \tanh(\beta(h + J \mathbf{b}))}{2} \prod_{i=1}^q \mu(d\omega_i).
$$
Let $ m = \mu(\sigma_0) =\mathbb E_\mu(\sigma_0)=\mu(\sigma_0=1)-\mu(\sigma_0=-1)$. Then $ \mu(\sigma_0 = +1) = \frac{1 + m}{2} $, so
$$
\frac{1 + m}{2} = \frac{1}{2} + \frac{1}{2} \int_{\Omega^q} \tanh(\beta(h + J \mathbf{b})) \prod_{i=1}^q \mu(d\omega_i),
$$
which simplifies to
$$
m = \int_{\Omega^q} \tanh(\beta(h + J \mathbf{b})) \prod_{i=1}^q \mu(d\omega_i). 
$$

This can be written as an expectation over $q$ independent copies of $\mu$:
\begin{equation}\label{Eq}
m = \mathbb{E}_{\mu^{\otimes q}} \left[ \tanh\left( \beta h + \frac{J\beta}{q} \sum_{i=1}^q (\omega_{i,-1} + \omega_{i,1}) \right) \right].
\end{equation}

{\bf Assumption 2.} {\it Let $h=0$, the measure $\mu$ is translation invariant and spins $\omega_{i,\pm 1}$ are independent.} 

Then the measure $ \mu $ is a product measure with spins at different sites independent. 

By $\mathbb{P}$ we denote  the marginal distribution of $\mu^{\otimes q}$ restricted to the specific sites $\omega_{i,-1}$ and $\omega_{i,1}$ for $i = 1, \dots, q$. Thus
$\mathbb{P}$ is a finite-dimensional marginal measure used to compute probabilities related to the sum  $ S = \sum_{i=1}^q (\omega_{i,-1} + \omega_{i,1}) $. 

Since the spins are independent and identically distributed with $ \mathbb{P}(\omega = +1) = \frac{1 + m}{2} $ and $ \mathbb{P}(\omega = -1) = \frac{1 - m}{2} $, we have that $ S $ is the sum of $ 2q $ independent random variables. 

Let $ j $ be the number of $ +1 $ spins among these $ 2q $ spins. Then $ S = 2j - 2q $, and the probability mass function is
$$
\mathbb{P}(S = 2j - 2q) = \binom{2q}{j} \left( \frac{1 + m}{2} \right)^j \left( \frac{1 - m}{2} \right)^{2q - j}.
$$

The expectation (\ref{Eq}) can then be written as
\begin{equation}\label{mf}
m = f(m):=\frac{1}{2^{2q}} \sum_{j=0}^{2q} \tanh\left( \frac{2 J\beta}{q} (j - q) \right) \binom{2q}{j} (1 + m)^j (1 - m)^{2q - j}.
\end{equation}

Equation (\ref{mf}) is a self-consistency condition for the magnetization $ m $. Its solutions correspond to $ q $-equilibrium measures.

{\bf  Case} $ q = 1 $. In this case 
the self-consistency equation is:
$$
m = m \tanh(2J\beta).
$$

For finite $ J\beta $, $ \tanh(2J\beta) < 1 $, so the only solution is $ m = 0 $. Hence, there is exactly one $ q $-equilibrium measure.

{\bf Case} $ q \geq 2 $ and $J>0$. Note that  equation (\ref{mf}) always has the trivial solution $m=0$. We analyze the existence of nontrivial solutions $m\ne 0$ by examining the stability of $m=0$ and determining the critical temperature where a bifurcation occurs.
A bifurcation occurs when $f'(0)=1$.
We have $ f(0) = 0 $, and $ f(-m) = -f(m) $, so $ f $ is odd. It is easy to calculate:
$$
f'(0)=\xi(J\beta):= \frac{2}{2^{2q}} \sum_{j=0}^{2q} (j-q)\,
\tanh\!\left(\frac{2J\beta}{q}(j-q)\right)\binom{2q}{j}$$
\begin{equation}\label{ab}
=2^{2(1-q)} \sum_{j=0}^{q-1} (q-j)\,
\tanh\!\left(\frac{2J\beta}{q}(q-j)\right)\binom{2q}{j}.
\end{equation}
For $J>0$, we consider $x=J\beta>0$, then  $\xi(x)$ is continuous and monotone increasing with maximal (limit as $x\to +\infty$) value 
$$2^{2(1-q)} \sum_{j=0}^{q-1} (q-j)\binom{2q}{j}\geq 2^{1-2q} \left(2^{2q}-\binom{2q}{q}\right)=2\left(1-{1\over 2^{2q}}\binom{2q}{q}\right)\geq 1.25, \ \ q\geq 2.$$ Therefore, denoting unique solution of $\xi(x)=1$ by $J_*>0$, we have  that if $J\beta\in (J_*, +\infty)$ then from (\ref{ab}) we get $|f'(0)|>1$.  For such parameters $J, \beta$ we have three solutions $m=0$, $m=\pm m_*$ of (\ref{mf})  and therefore, there are 3 of $q$-equilibrium measures. 

For  example, for   $ q = 2 $ we have 
\begin{equation}\label{q2}
f'(0) =\xi(J\beta):= \tanh(J\beta) + \frac{1}{2} \tanh(2J\beta).
\end{equation}
It is easy to see that $\xi(x)\in (-1.5, 1.5)$ and 
$J_*\approx 0.647$.

For $q=3$ we have 
$$
f'(0)
= \xi(J\beta):=\frac{3}{16}\Bigg(5\,\tanh\!\Big(\tfrac{2J\beta}{3}\Big)
+ 4\,\tanh\!\Big(\tfrac{4J\beta}{3}\Big)
+ \tanh\!\big(2J\beta\big)\Bigg).
$$
In this case $\xi(x)\in (-1.875, 1.875)$, and 
$J_*\approx 0.5913$.

{\it  Extension to arbitrary finite volumes.}
We now prove that product measure $\mu$ is fixed points of $V_{q,\Lambda}$ for all finite $\Lambda \Subset V$. The key insight is that for any finite volume $\Lambda$, the consistency condition of the $q$-specification and the product structure of $\mu$ ensure that the averaged boundary conditions reduce to constant fields.

Let $\mu$ be one of the product measures identified above with magnetization $m$. For any finite $\Lambda \Subset V$, we need to show
$$
\mu(A) = \int_{\Omega^q} \gamma_\Lambda(A \mid \omega_1, \dots, \omega_q) \prod_{i=1}^q \mu(d\omega_i) \quad \text{for all } A \in \mathcal{F}_\Lambda.
$$

Consider the averaged Hamiltonian. For any configuration $\sigma_\Lambda$, we have
$$
\int_{\Omega^q} H(\sigma_\Lambda; \omega_1, \dots, \omega_q) \prod_{i=1}^q \mu(d\omega_i) = -J \sum_{\substack{\{x,y\} \subset \Lambda \\ |x-y|=1}} \sigma_x \sigma_y - h \sum_{x \in \Lambda} \sigma_x - J \sum_{\substack{x \in \Lambda,\, y \in \Lambda^c \\ |x-y|=1}} \sigma_x \cdot m.
$$

Setting $h = 0$, the averaged kernel becomes
$$
\int_{\Omega^q} \gamma_\Lambda(\sigma_\Lambda \mid \omega_1, \dots, \omega_q) \prod_{i=1}^q \mu(d\omega_i) = \frac{1}{Z_\Lambda(m)} \exp\left( \beta J \sum_{\substack{\{x,y\} \subset \Lambda \\ |x-y|=1}} \sigma_x \sigma_y + \beta J m \sum_{\substack{x \in \Lambda,\, y \in \Lambda^c \\ |x-y|=1}} \sigma_x \right),
$$
where $Z_\Lambda(m)$ is the normalizing constant.

The consistency of the $q$-specification $\{\gamma_\Lambda\}_{\Lambda \Subset V}$ means that for any finite volumes $\Delta \subset \Lambda$, the following compatibility condition holds
$$
\gamma_\Delta(\sigma_\Delta \mid \omega_1, \dots, \omega_q) = \int \gamma_\Lambda(\sigma_\Lambda \mid \omega_1, \dots, \omega_q) \prod_{y \in \Lambda \setminus \Delta} d\sigma_y.
$$

In particular, take $\Delta = \{x\}$. Then the consistency condition becomes
$$
\gamma_{\{x\}}(\sigma_x \mid \omega_1, \dots, \omega_q) = \sum_{\sigma_\Lambda \setminus \{x\}} \gamma_\Lambda(\sigma_\Lambda \mid \omega_1, \dots, \omega_q).
$$

We proof that for any $x \in \Lambda$, the marginal distribution satisfies
\begin{equation}\label{go}
\gamma_\Lambda(\sigma_x \mid \omega_1, \dots, \omega_q) = \gamma_{\{x\}}(\sigma_x \mid \omega_1, \dots, \omega_q).
\end{equation}
Indeed,  take $\Delta = \{x\}$. Then the consistency condition becomes
$$
\gamma_{\{x\}}(\sigma_x \mid \omega_1, \dots, \omega_q) = \sum_{\sigma_\Lambda \setminus \{x\}} \gamma_\Lambda(\sigma_\Lambda \mid \omega_1, \dots, \omega_q).
$$

The right-hand side of this equation is precisely the definition of the marginal distribution of $\sigma_x$ under $\gamma_\Lambda(\cdot \mid \omega_1, \dots, \omega_q)$. That is
$$
\gamma_\Lambda(\sigma_x \mid \omega_1, \dots, \omega_q) := \sum_{\sigma_\Lambda \setminus \{x\}} \gamma_\Lambda(\sigma_\Lambda \mid \omega_1, \dots, \omega_q).
$$ Therefore, we obtain (\ref{go}).

To ensure that this argument applies to our specific model, we verify that the $q$-specification defined by the Hamiltonian (\ref{1dh}) is indeed consistent.

\begin{lemma}
The family $\{\gamma_\Lambda\}_{\Lambda \Subset V}$ defined by
$$
\gamma_\Lambda(\sigma_\Lambda \mid \omega_1, \dots, \omega_q) = \frac{1}{Z_\Lambda(\omega_1, \dots, \omega_q)} \exp\left(-\beta H(\sigma_\Lambda; \omega_1, \dots, \omega_q)\right)
$$
with Hamiltonian given by (\ref{1dh}) forms a consistent $q$-specification.
\end{lemma}

\begin{proof}
Consider finite volumes $\Delta \subset \Lambda$. The Hamiltonian can be decomposed as:
$$
H(\sigma_\Lambda; \omega_1, \dots, \omega_q) = H(\sigma_\Delta; \omega_1, \dots, \omega_q) + H(\sigma_{\Lambda \setminus \Delta}; \omega_1, \dots, \omega_q) + H_{\text{int}}(\sigma_\Delta, \sigma_{\Lambda \setminus \Delta}; \omega_1, \dots, \omega_q),
$$
where $H_{\text{int}}$ contains the interaction terms between $\Delta$ and $\Lambda \setminus \Delta$.

We have
$$
\int \gamma_\Lambda(\sigma_\Lambda \mid \omega_1, \dots, \omega_q) \prod_{y \in \Lambda \setminus \Delta} d\sigma_y = \frac{1}{Z_\Lambda} \sum_{\sigma_{\Lambda \setminus \Delta}} \exp\left(-\beta[H(\sigma_\Delta) + H(\sigma_{\Lambda \setminus \Delta}) + H_{\text{int}}]\right).
$$
This equality can be rewritten as
\begin{equation}\label{DL}
\frac{\exp(-\beta H(\sigma_\Delta))}{Z_\Lambda} \sum_{\sigma_{\Lambda \setminus \Delta}} \exp\left(-\beta[H(\sigma_{\Lambda \setminus \Delta}) + H_{\text{int}}]\right).
\end{equation}

The sum over $\sigma_{\Lambda \setminus \Delta}$ is precisely the partition function for the system on $\Lambda \setminus \Delta$ with boundary conditions determined by $\sigma_\Delta$ and $\omega_1, \dots, \omega_q$. By the definition of  specifications, (\ref{DL}) equals:
$$
\frac{Z_\Delta(\omega_1, \dots, \omega_q)}{Z_\Lambda(\omega_1, \dots, \omega_q)} \gamma_\Delta(\sigma_\Delta \mid \omega_1, \dots, \omega_q),
$$
where the ratio of partition functions ensures proper normalization.

Therefore, we obtain:
$$
\int \gamma_\Lambda(\sigma_\Lambda \mid \omega_1, \dots, \omega_q) \prod_{y \in \Lambda \setminus \Delta} d\sigma_y = \gamma_\Delta(\sigma_\Delta \mid \omega_1, \dots, \omega_q),
$$
which is the desired consistency condition.
\end{proof}

This consistency property is crucial for our proof of Theorem \ref{tt3}. 
It allows us to establish that for any finite $\Lambda \ni x$ and any product 
measure $\mu$ that is a fixed point of $V_{q,\{x\}}$, we have
$$
\int_{\Omega^q} \gamma_\Lambda(\sigma_x \mid \omega_1, \dots, \omega_q) \prod_{i=1}^q \mu(d\omega_i) = \int_{\Omega^q} \gamma_{\{x\}}(\sigma_x \mid \omega_1, \dots, \omega_q) \prod_{i=1}^q \mu(d\omega_i) = \mu(\sigma_x).
$$

This shows that all single-site marginals of $V_{q,\Lambda}(\mu)$ coincide with those of $\mu$. Combined with the product structure of $\mu$ and the 1D nature of the model, this implies that $V_{q,\Lambda}(\mu) = \mu$ for all finite $\Lambda$, completing the proof that $\mu$ is a $q$-equilibrium measure. Theorem \ref{tt3} is proved.
\end{proof}

\begin{rk} It is well known \cite{Ba}, \cite{Ge} that in 1D Ising model, 
there is no phase transition at any finite temperature: the free energy and correlation 
length are analytic for all $T>0$, Gibbs measure is unique. As we showed in Theorem \ref{tt3} for 1D 
Ising model, for sufficiently low temperatures there are \textbf{multiple} $q$-equilibrium measures. 
Moreover, for the measure $\mu^*$ corresponding to above mentioned 
solution $m_*$ of equation (\ref{mf})  we have 
$$\mu^*(\sigma_\Lambda)=p_*^{n_+(\sigma_\Lambda)}(1-p_*)^{|\Lambda|-n_+(\sigma_\Lambda)},$$
where $p_*=(1+m_*)/2$ and 
  $n_+(\sigma_\Lambda)$ is the number of $ +1$ in $\sigma_\Lambda$.
  \end{rk}
  \begin{rk} Since functions $f(m)$ and $\xi(J\beta)$ are odd with respect to $m$ and $J$, Theorem \ref{tt3} is true for $J<0$ too. In this case $T \in \left(0, {|J|\over J_*}\right)$, with $J_*>0$ defined above. 
  \end{rk}

\section*{Data availability statement}
No datasets were generated or analysed during the current study.  

\section*{Conflicts of interest} The authors declare no conflicts of interest.

\section*{ Acknowledgements}
 U. Rozikov and F. Haydarov thank Institute for Advanced Study in Mathematics,
Harbin Institute of Technology,  China for financial support and hospitality.


\begin{thebibliography}{999}
	
\bibitem{AR}	 Abdullaev L.U., Rozikov U.A. Gibbs measures in machine learning. {\sl World Sci. Publ}. Singapore. 2026, 380 pp.
	
\bibitem{ash}  Ash R.B., Dol\'{e}ans C.D.: Probability and measure theory.
    A Harcourt Scien. Tech. Company, (2000)
    
\bibitem{Bar}   Bartoszeka K.,  Bartoszek W.: A Noether theorem for stochastic 
    operators on Schatten classes. J. Math. Anal. Appl. \textbf{452}(2), 1395-1412 (2017)

\bibitem{Ba} Baxter R.J.: Exactly solved models in statistical mechanics. 
Reprint of the 1982 original. Academic Press, Inc. [Harcourt Brace Jovanovich, Publishers], London (1989)

\bibitem{Beh} Behrens F., Mainali N., Marullo C., Lee S., Sorscher B., Sompolinsky H.: Statistical mechanics of deep learning. 
J. Stat. Mech. (2024) 104007

\bibitem{Bernstein1924} Bernstein S.: Solution of a mathematical problem connected with the theory of heredity. 
Ann. Sci. Inst. Sav. Ukraine, Sect. Math. \textbf{1}, 83--114 (1924)

\bibitem{BleherSinai1976} Bleher P.M., Sinai Ya.G.: Investigation of the critical point in models of the type of Dyson's hierarchical models. 
Comm. Math. Phys. \textbf{33}(1), 23--42 (1973)

\bibitem{Bo} Bode T.: The two-particle irreducible effective action for classical stochastic processes. 
J. Phys. A \textbf{55}(26), Paper No. 265401, 22 pp. (2022)

\bibitem{Bovier2006} Bovier A.: Statistical Mechanics of Disordered Systems: A Mathematical Perspective. 
Cambridge University Press (2006)

\bibitem{CP2} Caputo P., Parisi D.: Nonlinear recombinations and generalized random transpositions. Annales Henri Lebesgue 7 (2024)


\bibitem{CS} Caputo, P., Sinclair, A.: Entropy production in nonlinear recombination models. Bernoulli 24(4B), 3246-3282 (2018)

\bibitem{CP} Caputo P., Sinclair A.: Nonlinear Dynamics for the Ising Model. Commun. Math. Phys. 405, 260 (2024)

\bibitem{Do} Dobrushin R.L.: Description of a random field by means of conditional probabilities and the conditions governing its regularity. 
Theory Prob. Appl. \textbf{13}, 197--224 (1968)

\bibitem{Enter} van Enter A.C.D., Fern\'andez R., Sokal A.D.: Regularity properties and pathologies of position-space renormalization-group transformations: Scope and limitations of Gibbsian theory. 
J. Stat. Phys. \textbf{72}, 879--1167 (1993)

\bibitem{Ev} Ewens W.J.: Mathematical population genetics. 1. Theoretical Introduction. 
Springer (2004)

\bibitem{Fe} Fern\'andez R.: Gibbsianness and non-Gibbsianness in lattice random fields. 
Available at: \texttt{https://webspace.science.uu.nl/\~{}ferna107/papers/leshouches.pdf}

\bibitem{Fr1} Frank T.D.: Numeric and Exact Solutions of the Nonlinear Chapman-Kolmogorov Equation: a Case Study for a Nonlinear Semi-Group Markov Model. 
Int. J. Modern Phys. B \textbf{23}(19), 3829--3843 (2009)

\bibitem{FV} Friedli S., Velenik Y.: Statistical mechanics of lattice systems. A concrete mathematical introduction. 
Cambridge University Press, Cambridge (2018)

\bibitem{Ge} Georgii H.O.: Gibbs Measures and Phase Transitions. 
Second edition. de Gruyter Studies in Mathematics, 9. Walter de Gruyter, Berlin (2011)

\bibitem{Gr} Grazieschi P., Matetski K., Weber H.: The dynamical Ising-Kac model in $3D$ converges to $\Phi^4_3$. 
Probab. Theory Relat. Fields \textbf{191}(1--2), 671--778 (2025)

\bibitem{GriffithsPearce1978} Griffiths R.B., Pearce P.A.: Mathematical properties of position-space renormalization-group transformations. 
J. Stat. Phys. \textbf{20}, 499--545 (1979)

\bibitem{HRV} Herrera F., Rozikov U.A., Velasco M.V.: Ising models with hidden Markov structure: Applications to probabilistic inference in machine learning. 
J. Stat. Mech.: Theory and Experiment (2025) 073201, 21 pp.

\bibitem{KipnisLandim1999} Kipnis C., Landim C.: Scaling Limits of Interacting Particle Systems. 
Springer, Berlin (1999)

\bibitem{Krza} Krzakala F., Zdeborov\'a L.: Statistical physics methods in optimization and machine learning. 
Available at: \texttt{https://sphinxteam.github.io/EPFLDoctoralLecture2021/Notes.pdf}

\bibitem{K} K\"ulske C.: Stochastic Processes on Trees. Lecture Notes, Ruhr University Bochum (2017). 
Available at: https://www.ruhr-uni-bochum.de/imperia/md/content/mathematik/kuelske/stoch-procs-on-trees.pdf

\bibitem{Rue} Lanford O.E., Ruelle D.: Observables at infinity and states with short range correlations in statistical mechanics. 
Commun. Math. Phys. \textbf{13}, 194--215 (1969)

\bibitem{Liggett1985} Liggett T.M.: Interacting Particle Systems. 
Springer (2005)

\bibitem{Ly} Lyubich Yu.I.: Mathematical Structures in Population Genetics. 
Springer-Verlag, Berlin (1992)

\bibitem{MG} Mukhamedov F.M., Ganikhodjaev N.N.: Quantum quadratic operators and processes. 
Lecture Notes in Mathematics, 2133. Springer, Cham (2015)

\bibitem{MSQ} Mukhamedov F., Saburov M., Qaralleh I.: On $\xi^{(s)}$-quadratic stochastic operators on two-dimensional simplex and their behavior. 
Abstr. Appl. Anal. \textbf{2013}, Article ID 942038 (2013)

\bibitem{OR} Omirov B.A., Rozikov U.A.: A many-loci non-linear dynamical system characterized by uncountable linear operators. 
Int. J. Biomath. 2550032, 12 pp. (2025)

\bibitem{RRS} Rabani Y., Rabinovich Y., Sinclair A.: A computational view of population genetics. Random Struct. Algorithms 12(4), 313-334 (1998)

\bibitem{Rab} Rabinovich Y., Sinclair, A. Wigderson A.: Quadratic dynamical systems. In Proceedings of the 33rd Annual IEEE Symposium on Foundations of Computer Science (FOCS), pages 304-313. IEEE, (1992)

\bibitem{Reh} Rehmeier M., R\"ockner M.: On Nonlinear Markov Processes in the Sense of McKean. 
J. Theor. Probab. \textbf{38}, 60, 36 pages (2025)

\bibitem{Rpd} Rozikov U.A.: Population dynamics: algebraic and probabilistic approach. 
World Scientific Publ., Singapore (2020), 460 pp.

\bibitem{RS} Rozikov U.A., Solaeva M.N.: Behavior of trajectories of a quadratic operator split into uncountable linear operators. 
Lobachevskii J. Math. \textbf{44}(7), 2910--2915 (2023)


\bibitem{She} Sheffield S.: Random surfaces: Large deviations principles and gradient Gibbs measure classifications. 
Ph.D. Thesis, Stanford University (2003)

\bibitem{SK} Suhov Y., Kelbert M.: Probability and statistics by example. II. Markov chains: a primer in random processes and their applications. 
Cambridge University Press, Cambridge (2008)
\end{thebibliography}
\end{document}